\documentclass[aps,prd,preprintnumbers,groupedaddress,nofootinbib,amssymb,eqsecnum,notitlepage]{revtex4}
\usepackage{here}
\usepackage{graphicx}
\usepackage{amsmath,amsthm,amssymb}
\usepackage{bm}
\usepackage{color}

\usepackage{amsfonts}

\usepackage{dcolumn}
\usepackage{hyperref}
\allowdisplaybreaks[1]

\begin{document}
\newcommand{\newc}{\newcommand}

\newc{\rk}[1]{{\color{red} #1}}
\newc{\ben}{\begin{eqnarray}}
\newc{\een}{\end{eqnarray}}
\newc{\be}{\begin{equation}}
\newc{\ee}{\end{equation}}
\newc{\ba}{\begin{eqnarray}}
\newc{\ea}{\end{eqnarray}}
\newc{\D}{\partial}
\newc{\rH}{{\rm H}}
\newc{\dphi}{\delta\phi}
\newc{\pa}{\partial}
\newc{\tp}{\dot{\phi}}
\newc{\ttp}{\ddot{\phi}}
\newc{\drhoc}{\delta\rho_c}
\newc{\aB}{\alpha_{\rm B}}
\newc{\aK}{\alpha_{\rm K}}
\newc{\aM}{\alpha_{\rm M}}
\newc{\bn}{\beta_{n_c}}
\newc{\bK}{\beta_{\rm K}}
\newc{\delc}{\delta_{c{\rm N}}}
\newc{\eH}{\epsilon_{\rm H}}
\newc{\ep}{\epsilon_{\phi}}

\preprint{WUCG-20-03}

\title{General formulation of cosmological perturbations in 
scalar-tensor dark energy coupled to dark matter}

\author{Ryotaro Kase$^{1}$ and Shinji Tsujikawa$^{2}$}

\affiliation{
$^1$Department of Physics, Faculty of Science, 
Tokyo University of Science, 1-3, Kagurazaka,
Shinjuku-ku, Tokyo 162-8601, Japan\\
$^2$Department of Physics, Waseda University, 3-4-1 Okubo, Shinjuku, Tokyo 169-8555, Japan}

\begin{abstract}

For a scalar field $\phi$ coupled to cold dark matter (CDM), 
we provide a general framework for studying the background and perturbation dynamics on 
the isotropic cosmological background. The dark energy sector is described by a Horndeski 
Lagrangian with the speed of gravitational waves equivalent to that of light, whereas
CDM is dealt as a perfect fluid characterized by the number density $n_c$ and four-velocity $u_c^\mu$.
For a very general interacting Lagrangian $f(n_c, \phi, X, Z)$, where $f$ depends on 
$n_c$, $\phi$, $X=-\partial^{\mu} \phi \partial_{\mu} \phi/2$, 
and $Z=u_c^{\mu} \partial_{\mu} \phi$, 
we derive the full linear perturbation equations of motion without fixing any gauge conditions. 
To realize a vanishing CDM sound speed for the successful structure formation, 
the interacting function needs to be of the form $f=-f_1(\phi, X, Z)n_c+f_2(\phi, X, Z)$. 
Employing a quasi-static approximation for the modes deep inside the sound horizon, 
we obtain analytic formulas for the effective gravitational couplings of CDM and 
baryon density perturbations as well as gravitational and weak lensing potentials.
We apply our general formulas to several interacting theories and show that, 
in many cases, the CDM gravitational coupling around the quasi de-Sitter background 
can be smaller than the Newton constant $G$ due to a momentum transfer induced by 
the $Z$-dependence in $f_2$.

\end{abstract}

\date{\today}


\maketitle

\section{Introduction}
\label{introsec}

Today's universe is dominated by two unknown components--dark energy (DE)
and dark matter (DM).  The late-time cosmic acceleration is driven by 
a negative pressure of DE, whereas DM is the main source for 
gravitational clusterings.
The standard cosmological paradigm is known as the 
$\Lambda$-cold-dark-matter ($\Lambda$CDM) 
model \cite{Peebles1,Peebles2}, in which 
the origin of DE is the cosmological constant with 
nonrelativistic DM fluids.  
The cosmological constant has no dynamical propagating degrees 
of freedom, so it does not couple to CDM directly. 
On the theoretical side, it is generally difficult to reconcile the observed 
energy scale of $\Lambda$ with the vacuum energy arising from
particle physics \cite{Weinberg:1988cp}. 
In recent observations, the $\Lambda$CDM model is plagued by tensions of 
today's expansion rate $H_0$ as well as the amplitude of matter density contrast $\sigma_8$
between the Cosmic-Microwave-Background (CMB) and low-redshift 
measurements \cite{Planck2015,Verde:2019ivm,Riess:2019cxk,Freedman:2019jwv,weak1,weak2}.

The cosmological constant is not the only possibility for the origin of DE, 
but there are other dynamical models of DE proposed in the literature 
(see Refs.~\cite{CST,Clifton,Joyce,Ishak:2018his,Kase:2018aps} for reviews).
The simple example of dynamical DE models is a scalar field $\phi$ with associated potential and 
kinetic energies \cite{quin1,quin2,quin3,quin4,quin5,quin6,quin7,kes1,kes2,kes3}.  
Theories with a single propagating scalar degree of freedom include Horndeski 
gravity \cite{Horndeski,Def11,KYY,Char11} 
and its extension to theories containing derivatives higher than second order--such as 
GLPV \cite{GLPV} and 
DHOST theories \cite{Langlois1,Langlois2,CKT}. 
In Horndeski theories the field equations of motion are strictly of second order on 
any curved background. 
After the gravitational-wave event GW170817 \cite{GW170817} together 
with the electromagnetic counterpart \cite{Goldstein}, the speed of tensor perturbations 
$c_t$ is constrained to be very close to that of light $c$ for the redshift $z<0.009$. 
If we strictly demand that $c_t=c$, the allowed Horndeski Lagrangian is of the form 
${\cal L}_{\rm H}=G_4(\phi)R+G_2(\phi,X)+G_3(\phi,X) 
\square \phi$ \cite{Lon15,GWcon1,GWcon2,GWcon3,GWcon4,GWcon5,GWcon6,Kase:2018iwp}, 
where $G_4$ depends only on $\phi$, and $G_{2,3}$ are functions of 
$\phi$ and $X=-\partial^{\mu} \phi \partial_{\mu} \phi/2$.

In this sub-class of Horndeski theories where the field $\phi$ is uncoupled to CDM, 
the gravitational coupling of CDM and baryon perturbations relevant to 
the scale of galaxy clusterings is usually larger than the Newton 
constant $G$ \cite{Tsujikawa:2015mga,GWcon5,Kase:2018aps}. 
This property is attributed to the fact that the scalar-matter interaction 
is attractive under the absence of ghost and Laplacian instabilities 
for perturbations deep inside the sound horizon. 
Then the growth of CDM and baryon density perturbations is enhanced with 
respect to the $\Lambda$CDM model, so the discrepancy of $\sigma_8$ 
between CMB and low-redshift measurements tends to be even worse. 
This means that, even with the cosmological background reducing the 
tension of $H_0$ in comparison to the $\Lambda$CDM, the 
problem of $\sigma_8$ discrepancy still persists 
at the level of perturbations \cite{Alexander}.

If the scalar field is coupled to CDM, there are intriguing possibilities that the gravitational couplings 
for CDM are smaller than $G$ for linear perturbations associated with the growth 
of large-scale structures. In particular, the derivative interaction between the CDM 
four-velocity $u_c^{\mu}$ and the field derivative $\partial_{\mu} \phi$, 
which is weighed by the scalar product $Z=u_c^{\mu} \partial_{\mu} \phi$, 
allows the possibility of weak cosmic growth through a momentum 
transfer \cite{Pourtsidou:2013nha,Boehmer:2015kta,Boehmer:2015sha,Skordis:2015yra,
Koivisto:2015qua,Pourtsidou:2016ico,Dutta:2017kch,Linton,Kase:2019veo,Kase:2019mox,
Chamings:2019kcl,Amendola:2020ldb,Jimenez:2020ysu}. 
Besides this momentum transfer, the energy exchange between CDM 
and the scalar field can be also accommodated by 
implementing the CDM number density $n_c$ coupled to 
$\phi$ and $X$ \cite{Pourtsidou:2013nha,Boehmer:2015kta,Kase:2019veo,Amendola:2020ldb}. 
This dependence of $n_c$ in the interacting Lagrangian can be interpreted as 
the dependence of CDM density $\rho_c=m_c n_c$, where $m_c$ is the mass 
of CDM particles.
These coupled DE and DM theories in the Lagrangian formulation 
have a theoretical advantage over the phenomenological approaches taken in 
Refs.~\cite{Dalal:2001dt,Zimdahl:2001ar,Chimento:2003iea,Wang1,Wei:2006ut,Amendola:2006dg,Guo:2007zk,
Valiviita:2008iv,Gavela:2009cy,Wands,Kumar:2016zpg,DiValentino:2017iww,An:2017crg,
Yang:2018euj,Pan:2019gop,DiValentino:2019ffd,Yang:2019uog,DiValentino:2019jae,Vagnozzi:2019kvw}, in that the background and 
perturbation equations of motion in former theories unambiguously follow from the 
interacting Lagrangian \cite{Tamanini:2015iia}.

The interacting Lagrangian containing the effect of both energy and momentum 
transfers is generally expressed in the form $f(n_c, \phi, X, Z)$, 
where $f$ is a function of $n_c$, $\phi$, $X$, and $Z$. 
So far, the background and perturbation dynamics of such interacting theories 
have been studied for several sub-classes of the coupling $f(n_c, \phi, X, Z)$, 
with the DE sector restricting to
quintessence \cite{Pourtsidou:2013nha,Boehmer:2015kta,Boehmer:2015sha,Skordis:2015yra,
Koivisto:2015qua,Pourtsidou:2016ico,Dutta:2017kch,Linton}. 
In Ref.~\cite{Kase:2019veo}, the present authors considered the 
Lagrangian of Horndeski theories for the DE scalar field, 
but the interacting function is restricted to be of the form 
$f=-f_1(\phi,X)n_c+f_2(n_c,\phi,X)Z$. 
In Ref.~\cite{Amendola:2020ldb}, the dynamics of perturbations for an interacting Lagrangian
$f=-f_1(\phi,X,Z)n_c+f_2(\phi,X,Z)$ was studied by choosing specific functional 
forms of $f_1$ and $f_2$.

In this paper, we study the cosmology of coupled DE and DM theories characterized by 
the interacting function $f(n_c, \phi, X, Z)$, with the aforementioned Horndeski 
Lagrangian ${\cal L}_{\rm H}$ in the DE sector. 
The CDM is dealt as a perfect fluid, which is described by a Schutz-Sorkin 
action \cite{Sorkin,Brown,DGS}. 
We obtain the field equations of motion in a covariant form and apply them 
to the flat Friedmann-Lema\^{i}tre-Robertson-Walker (FLRW) background.
We then expand the action up to second-order in scalar perturbations without fixing 
gauge conditions and derive the full linear perturbation equations of motion in a gauge-ready form.
We also identify conditions for the absence of ghost and Laplacian instabilities and show 
that the condition $\partial^2 f/\partial n_c^2=0$ must be satisfied to realize the 
vanishing effective CDM sound speed. 
In this class of theories, we obtain the effective gravitational couplings of CDM and baryons 
by employing the quasi-static approximation for perturbations deep inside the sound horizon. 
Finally, we apply our general formulas to several concrete theories of coupled DE and DM 
and show that the weak cosmic growth around the quasi de Sitter background
is possible in many cases by the momentum exchange.

Throughout the paper, we use the natural units for which the speed of 
light $c$, the reduced Planck constant $\hbar$, and the
Boltzmann constant $k_B$ are set to unity. 
We also adopt the metric signature $(-,+,+,+)$, with the Greek and Latin indices 
representing components in four-dimensional space-time and in 
three-dimensional space, respectively.

\section{Horndeski scalar coupled to DM}
\label{eomsec}

We study a general Lagrangian formulation of coupled DE in which 
a scalar field $\phi$ interacts with CDM 
through both energy and momentum exchanges. 
We assume that the scalar field couples to neither baryons 
nor radiation. For the DE sector, we consider a sub-class of Horndeski 
theories in which the speed of gravitational waves is identical to that 
of light \cite{Lon15,GWcon1,GWcon2,GWcon3,GWcon4,GWcon5,GWcon6,Kase:2018iwp}. 
We deal with CDM, baryons, and radiation as perfect fluids.

The total action is given by 
\be
{\cal S} =
\int {\rm d}^4 x \sqrt{-g}\,{\cal L}_{\rm H}
-\sum_{I=c,b,r}\int {\rm d}^{4}x 
\left[\sqrt{-g}\,\rho_I(n_I)+ J_I^{\mu} \partial_{\mu} \ell_I \right]
+\int {\rm d}^4x\sqrt{-g}\,f(n_c, \phi, X, Z)\,,
\label{action}
\ee
where 
\be
{\cal L}_{\rm H}=G_4(\phi)R+G_2(\phi,X)+G_3(\phi,X) \square \phi\,.
\label{LH}
\ee
Here, $g$ is the determinant of metric tensor $g_{\mu\nu}$, 
$R$ is the Ricci scalar, $G_4$ is a function of $\phi$,
and $G_{2,3}$ depend on both $\phi$ and its kinetic term
\be
X=-\frac{1}{2}\nabla^{\mu}\phi \nabla_{\mu} \phi\,,
\ee
with the covariant derivative operator $\nabla_{\mu}$ 
and the d'Alembertian 
$\square \equiv g^{\mu \nu} \nabla_{\mu} \nabla_{\nu}$. 

The second integral on the right-hand-side of Eq.~(\ref{action}), 
which corresponds to the Schutz-Sorkin action \cite{Sorkin,Brown,DGS},  
describes the perfect fluids of CDM, baryons, and 
radiation (labeled by $c,b,r$, respectively). 
The energy density $\rho_I$ is a function of 
each fluid number density $n_I$. 
The vector field $J_I^{\mu}$ in the Schutz-Sorkin action
is related to the fluid four-velocity $u_I^{{\mu}}$, as
\be
u_I^{{\mu}}=\frac{J_I^{{\mu}}}{n_I\sqrt{-g}}\,.
\label{udef}
\ee
Since the four-velocity obeys $u_I^{\mu} u_{I{\mu}}=-1$, 
the explicit relation between $n_I$ and $J_I^{\mu}$ 
is given by 
\be
n_I=\sqrt{\frac{J_I^{\mu} J_{I\mu}}{g}}\,.
\label{ndef}
\ee
The scalar quantity $\ell_I$ in the Schutz-Sorkin action 
is a Lagrange multiplier, 
with the notation $\partial_{\mu} \ell_I=
\partial \ell/\partial x^{\mu}$.
Since the product $J_I^{\mu} \partial_{\mu} \ell_I$ is not multiplied 
by the volume factor $\sqrt{-g}$ in the action, 
we do not deal with $J_I^{\mu}$ as a covariant four vector.
Alternatively one can introduce the four vector 
$\tilde{J}_I^{\mu}=J_I^{\mu}/\sqrt{-g}$ and consider the covariant 
action $-\int {\rm d}^{4}x \sqrt{-g}\tilde{J}_I^{\mu} \nabla_{\mu} \ell_I$ \cite{SorkinADF}, 
but we do not take this latter approach in this paper.
There are also additional contributions associated with the 
dynamical vector degrees of freedom to the above Schutz-Sorkin 
action \cite{Sorkin,Brown,DGS}. 
Since vector perturbations are non-dynamical 
in scalar-tensor theories, we do not take them into account.

The third integral on the right-hand-side of Eq.~(\ref{action}) represents 
the interaction between the scalar field and CDM \cite{Pourtsidou:2013nha}.
The function $f$ depends on $n_c$, $\phi$, $X$, and 
\be
Z= u_c^{\mu}\nabla_{\mu}\phi
=\frac{J_c^{{\mu}}}{n_c\sqrt{-g}}\nabla_{\mu}\phi\,.
\ee
If we consider CDM with mass $m_c$, the corresponding density 
is given by $\rho_c=m_c n_c$ and hence 
$n_c$ is directly related to its density $\rho_c$. 
Then the $n_c$ dependence in $f$, along with the coupling with
$\phi$ and $X$, accommodates the energy transfer between 
CDM and DE. 
The $Z$ dependence characterizes a derivative interaction 
between the CDM four-velocity and the scalar derivative 
$\nabla_{\mu}\phi$, which mediates the momentum exchange. 
The scalar products $X$ and $Z$ are constructed by the 
first derivative of $\phi$. 
One can also consider more general 
scalar products containing the derivatives higher than 
the first order, but we will not do so in this paper to 
keep the equations of motion up to second order.

The interacting Lagrangian $f(n_c, \phi, X, Z)$ may be related to CDM 
conformally and disformally coupled to the metric $\bar{g}_{\mu \nu}=A(\phi, X) g_{\mu \nu}
+B(\phi,X) \nabla_{\mu} \phi \nabla_{\nu} \phi$ \cite{Gleyzes:2015pma,Kimura:2017fnq,Chibana:2019jrf}.
Under the disformal transformation, Horndeski theories are mapped
to GLPV or DHOST theories (see, e.g., Ref.~\cite{Langlois:2018dxi}). 
It may be possible to seek for the frame in which CDM is minimally coupled, i.e., 
the Jordan frame for CDM. 
In doing so, we need to caution that the product $J_I^{\mu}\partial_{\mu}\ell_I$ 
in the action (\ref{action}), which is associated with the conservation of total 
particle number for each fluid on the cosmological background, is subject to 
modifications under the disformal transformation.
It is of interest to study how the theories described by Eq.~(\ref{action}) 
can be expressed in such a transformed frame, but this is beyond the scope of our paper. 
We focus on the frame where the theories are given by the action (\ref{action}), 
i.e., the frame in which the total particle number of each fluid is conserved. 

\subsection{Covariant equations of motion}

We derive the covariant equations of motion for the theories 
given by Eq.~(\ref{action}). 
First, we vary the action (\ref{action}) with respect to 
the Lagrangian multiplier $\ell_I$ and obtain the following 
constraint,  
\be
\partial_{\mu}J_I^{\mu}=0\qquad
({\rm for}~I=c, b,r)\,.
\label{Jmu}
\ee
In terms of the four velocity $u_I^{\mu}$, 
this equation translates to 
$\partial_{\mu}(\sqrt{-g}\,n_Iu_I^{\mu})=0$. On using the relation 
$\partial_{\mu}(\sqrt{-g}u_I^{\mu})
=\sqrt{-g}\nabla_{\mu}u_I^{\mu}$, 
it follows that  
\be
u_I^{\mu}\partial_{\mu}n_I+n_I\nabla_{\mu}u_I^{\mu}=0\,.
\label{Jmu2}
\ee
Since the density $\rho_I$ depends on $n_I$ alone,  
it is straightforward to relate $\partial_{\mu}n_I$ 
with the partial derivative $\partial_{\mu}\rho_I$, 
such that 
\be
\rho_{I,n_I}u_I^{\mu}\partial_{\mu}n_I
=u_I^{\mu}\partial_{\mu}\rho_I\,,
\label{udrho}
\ee
where $\rho_{I,n_I}=\partial \rho_I/\partial n_I$. 
Then, we can express Eq.~(\ref{Jmu2}) in the form, 
\be
u_I^{\mu}\partial_{\mu}\rho_I
+(\rho_I+P_I)\nabla_{\mu}u_I^{\mu}=0\,, 
\label{umu}
\ee
where $P_I$ is the fluid pressure defined by 
\be
P_I=n_I\rho_{I,n_I}-\rho_I\,. 
\label{Pdef}
\ee
As we will see below, Eq.~(\ref{umu}) corresponds to the 
conservation (continuity) equation for the fluid 
energy-momentum tensor. 

Second, we vary the action (\ref{action}) with respect to the 
vector fields $J_I^{\mu}$. 
Taking note that the $J_c^{\mu}$ dependence in the action (\ref{action}) 
appears through
$\rho_c (n_c)$ as well as the $n_c$ and $Z$ dependence in $f$,
the variation with respect to 
the CDM vector field $J_c^{\mu}$ leads to 
\be
-\sqrt{-g} \rho_{c,n_c} \frac{\partial n_c}{\partial J_c^{\mu}}
- \partial_{\mu} \ell_c+\sqrt{-g} \left( f_{,n_c} 
\frac{\partial n_c}{\partial J_c^{\mu}}+
f_{,Z} \frac{\partial Z}{\partial J_c^{\mu}} \right)=0\,,
\ee
where 
\ba
& &
\frac{\partial n_c}{\partial J_c^{\mu}}=
\frac{J_{c \mu}}{n_c g}=-\frac{u_{c \mu}}{\sqrt{-g}}\,,\\
& &
\frac{\partial Z}{\partial J_c^{\mu}}
=\frac{1}{n_c \sqrt{-g}} \left( \nabla_{\mu} \phi
+Z u_{c \mu} \right)\,.
\ea
Then, we obtain
\be
\partial_{\mu} \ell_{c}= 
\left(\rho_{c,n_c}-f_{,n_c} \right)u_{c{\mu}}+\frac{f_{,Z}}{n_c}
\left(\nabla_{\mu}\phi+Zu_{c\mu} \right) \,.
\label{ellc}
\ee
For baryons and radiation, the relations analogous to 
Eq.~(\ref{ellc}) are
\be
\partial_{\mu} \ell_{I}
= \rho_{I,n_I} u_{I{\mu}} \qquad({\rm for}~I=b,r)\,.
\label{ellbr}
\ee
We will exploit Eqs.~(\ref{ellc}) and (\ref{ellbr}) to eliminate 
the Lagrange multipliers $\ell_I$ from 
the covariant equations of motion. 

Third, we vary the action (\ref{action}) with respect to 
$g^{\mu\nu}$ to derive the gravitational equations 
of motion. The variation of the Horndeski Lagrangian 
$L_{\rm H}=\sqrt{-g} {\cal L}_{\rm H}$
is given by \cite{KYY}
\be
\frac{2}{\sqrt{-g}} \frac{\delta L_{\rm H}}{\delta g^{\mu \nu}}
=M_{\rm pl}^2 G_{\mu \nu}-T^{({\rm H})}_{\mu\nu}\,,
\label{LHva}
\ee
where $G_{\mu \nu}$ is the Einstein tensor, 
$M_{\rm pl}$ is the reduced Planck mass, and 
\ba
T^{({\rm H})}_{\mu\nu}&=&
G_2g_{\mu\nu}+G_{2,X}\nabla_{\mu}\phi\nabla_{\nu}\phi
+G_{3,X} \nabla_{\mu}\phi\nabla_{\nu}\phi\,\square \phi 
-g_{\mu \nu} \nabla_{\lambda} G_3 \nabla^{\lambda}\phi
+\nabla_{\mu} G_3 \nabla_{\nu} \phi
+\nabla_{\nu} G_3 \nabla_{\mu} \phi
\notag\\
&&
+(M_{\rm pl}^2-2G_4)G_{\mu \nu}
+2G_{4,\phi}\left(\nabla_{\mu}\nabla_{\nu}\phi-g_{\mu\nu}\Box\phi\right)
+2G_{4,\phi\phi}\left(\nabla_{\mu}\phi\nabla_{\nu}\phi
+2Xg_{\mu\nu}\right)\,.
\ea
In Eq.~(\ref{LHva}), we have separated the term 
$M_{\rm pl}^2 G_{\mu \nu}$ from the other contributions.
 
For the second and third integrals in the action (\ref{action}), 
we express them in the form 
${\cal S}_{\rm F}=\int {\rm d}^4 x\,L_{\rm F}$, 
where
\be
L_{\rm F}=-\sum_{I=c,b,r}\left[\sqrt{-g}\,\rho_I(n_I)+ J_I^{\mu} \partial_{\mu} \ell_I \right]
+\sqrt{-g}\,f(n_c, \phi, X, Z)\,.
\label{LF}
\ee
Variation of this Lagrangian with respect to $g_{\mu \nu}$ leads to 
\be
\delta L_{\rm F}=-\sum_{I=c,b,r}\left[\delta \sqrt{-g}\,\rho_I
+\sqrt{-g}\,\rho_{I,n_I} \delta n_I+ J_{I\mu} \partial_{\nu} \ell_I \delta g^{\mu \nu} \right]
+\delta \sqrt{-g}\,f+\sqrt{-g} \left( f_{,n_c} \delta n_c
+f_{,X}\delta X+f_{,Z}\delta Z \right)\,,
\ee
where 
\ba
\delta \sqrt{-g} 
&=& -\frac{1}{2} \sqrt{-g}\,
g_{\mu \nu} \delta g^{\mu \nu}\,,\\
\delta n_I &=& \frac{n_I}{2} \left( g_{\mu \nu} 
-u_{I\mu} u_{I\nu} \right) \delta g^{\mu \nu}\,,\\
\delta X &=&-\frac{1}{2} \nabla_{\mu} \phi
\nabla_{\nu} \phi\,\delta g^{\mu \nu}\,,\\
\delta Z &=& \left( \frac{1}{2} Z u_{c\mu} 
u_{c\nu} +u_{c\mu}\nabla_{\nu} \phi \right) 
\delta g^{\mu \nu}\,.
\ea
On using the properties (\ref{ellc}) and (\ref{ellbr}), 
it follows that 
\be
-\frac{2}{\sqrt{-g}} \frac{\delta L_{\rm F}}{\delta g^{\mu \nu}}
=\sum_{I=c,b,r}T^{(I)}_{\mu\nu}
+T^{({\rm int})}_{\mu\nu}\,,
\label{Lfva}
\ee
where 
\ba
T^{(I)}_{\mu\nu}&=&(\rho_I+P_I)u_{I{\mu}}u_{I{\nu}}+P_Ig_{\mu\nu}\,,
\label{TI}\\
T^{({\rm int})}_{\mu\nu}&=&
f g_{\mu\nu}-n_cf_{,n_c} \left( g_{\mu \nu}+u_{c\mu}u_{c\nu} \right)
+f_{,X}\nabla_{\mu}\phi\nabla_{\nu}\phi+Zf_{,Z}u_{c\mu}u_{c\nu}\,.
\ea
{}From Eqs.~(\ref{LHva}) and (\ref{Lfva}), the gravitational equations 
of motion are given by  
\be
M_{\rm pl}^2G_{\mu\nu}=
T^{({\rm H})}_{\mu\nu}
+\sum_{I=c,b,r}T^{(I)}_{\mu\nu}
+T^{({\rm int})}_{\mu\nu}\,.
\label{ein}
\ee
The energy-momentum tensors $T^{({\rm H})}_{\mu\nu}$, 
$T^{(I)}_{\mu\nu}$, and $T^{({\rm int})}_{\mu\nu}$ correspond 
to those arising from the Horndeski sector, perfect fluids, and 
the coupling $f$, respectively.
Taking the covariant derivative of Eq.~(\ref{ein}), 
we obtain
\be
\nabla^{\mu}T^{({\rm H})}_{\mu\nu}
+\sum_{I=c,b,r} \nabla^{\mu}T^{(I)}_{\mu\nu}
+\nabla^{\mu}T^{({\rm int})}_{\mu\nu}=0\,.
\label{conco}
\ee
On using Eq.~(\ref{umu}), the perfect-fluid 
energy-momentum tensor $T^{(I)}_{\mu\nu}$ obeys
\be
u_I^{\nu} \nabla^{\mu}T^{(I)}_{\mu\nu}
=-\left[ u_I^{\mu}\partial_{\mu}\rho_I
+(\rho_I+P_I)\nabla_{\mu}u_I^{\mu} \right]=0\,,
\label{ucov}
\ee
which is equivalent to the continuity 
equation (\ref{Jmu}).
If the four-velocities of CDM, baryons, and radiation 
are identical to each other (which is the case for the 
FLRW background) or there is only one fluid component characterized 
by the four-velocity $u^{\nu}$, then the continuity equation 
$u^{\nu} \nabla^{\mu}T^{(I)}_{\mu\nu}=0$ holds 
for each fluid or a single fluid.
In this case, Eq.~(\ref{conco}) gives 
\be
u^{\nu} \nabla^{\mu} \left( T^{({\rm H})}_{\mu\nu}
+T^{({\rm int})}_{\mu\nu} \right)=0\,,
\label{conphi}
\ee
which corresponds to the continuity equation 
for the scalar field.

We note that the function $f$ in $T^{({\rm int})}_{\mu\nu}$ 
contains the dependence of CDM density $\rho_c$ 
through the number density $n_c$. 
It is then possible to absorb such $\rho_c$-dependent terms 
into the standard CDM energy-momentum tensor 
$T_{\mu \nu}^{(c)}$. By defining the modified CDM 
energy-momentum tensor $\hat{T}_{\mu \nu}^{(c)}$ in this way, 
the continuity equations of CDM and scalar field possess 
explicit interacting terms associated with the energy 
transfer \cite{Kase:2019veo}.
In Sec.~\ref{bacsec}, we will explicitly see this for the coupling $f$ 
separable into $n_c$ and other variables.

\subsection{Background equations of motion} 
\label{bacsec}

We derive the background equations of motion 
on the flat FLRW background described by the line element
\be
{\rm d}s^2=-{\rm d}t^2
+a^2(t) \delta_{ij} {\rm d}x^i {\rm d}x^j\,,
\label{BGmet}
\ee
where $a(t)$ is the time-dependent scale factor.
On this background, the scalar field $\phi$ depends only 
on $t$. Each perfect fluid in the rest frame has the four-velocity
$u_I^{\mu}=(1,0,0,0)$, with $I=c,b,r$.
{}From Eq.~(\ref{udef}), the temporal component of $J_I^{\mu}$ 
is equivalent to $J_I^0 \equiv {\cal N}_I=n_I a^3$. 
Due to the constraint (\ref{Jmu}), we have 
\be
{\cal N}_I={\rm constant}\,,
\ee
which corresponds to the conservation of total particle number 
of each fluid. 
This relation is equivalent to the continuity Eq.~(\ref{umu}).
On the background (\ref{BGmet}), Eq.~(\ref{umu}) reduces to
\be
\dot{\rho}_I+3H \left( \rho_I+P_I \right)=0\,,\qquad {\rm for} \quad
I=c,b,r,
\label{coneq}
\ee
where a dot represents the derivative with respect to $t$, and 
$H=\dot{a}/a$ is the Hubble-Lema\^{i}tre expansion rate.

The (00) and $(ii)$ components of the gravitational Eq.~(\ref{ein}) are
given, respectively, by 
\ba
& &
3M_{\rm pl}^2 H^2=\rho_{\rm DE}+\sum_{I=c,b,r}\rho_I\,,\label{back1}\\
& &
M_{\rm pl}^2 \left( 2\dot{H}+3H^2 \right)
=-P_{\rm DE}-\sum_{I=c,b,r}P_I\,,
\label{back2}
\ea
where 
\ba
\hspace{-0.8cm}
\rho_{\rm DE} &=& -G_2+\dot{\phi}^2 G_{2,X}+\dot{\phi}^2 
\left( G_{3,\phi} -3H \dot{\phi} G_{3,X} \right)
+3H^2 \left( M_{\rm pl}^2-2G_4 \right)-6H \dot{\phi}G_{4,\phi}
-f+\dot{\phi}^2 f_{,X}+\dot{\phi}f_{,Z}\,,\\
\hspace{-0.8cm}
P_{\rm DE} &=& G_2+\dot{\phi}^2 \left( G_{3,\phi}+
\ddot{\phi} G_{3,X} \right)
+ \left( 2\dot{H}+3H^2 \right) \left( 2G_4-M_{\rm pl}^2 \right)
+2 \left( \ddot{\phi}+2H \dot{\phi} \right) G_{4,\phi}
+2\dot{\phi}^2 G_{4,\phi \phi}+f-n_c f_{,n_c}\,.
\ea
Defining the density parameters, 
\be
\Omega_{\rm DE}=\frac{\rho_{\rm DE}}{3M_{\rm pl}^2 H^2}\,,
\qquad 
\Omega_{I}=\frac{\rho_{I}}{3M_{\rm pl}^2 H^2}\,,
\label{Omedef}
\ee
the Hamiltonian constraint (\ref{back1}) is expressed in the form,
\be
\Omega_{\rm DE}+\sum_{I=c,b,r} \Omega_I=1\,.
\label{Omecon}
\ee
Taking the time derivative of Eq.~(\ref{back1}) and using 
Eqs.~(\ref{coneq}) and (\ref{back2}), we obtain 
\be
\dot{\rho}_{\rm DE}+3H \left( \rho_{\rm DE}+P_{\rm DE} 
\right)=0\,,
\label{coneq2}
\ee
which corresponds to Eq.~(\ref{conphi}). 
We define the equations of state of dark energy and 
perfect fluids, as
\be
w_{\rm DE}=\frac{P_{\rm DE}}{\rho_{\rm DE}}\,,\qquad
w_I=\frac{P_I}{\rho_I}\,.
\ee
For given $w_I$ ($I=c,b,r$) and initial conditions, the background 
dynamics is determined by integrating Eqs.~(\ref{coneq}), 
(\ref{back2}) and (\ref{coneq2}) 
together with the constraint Eq.~(\ref{Omecon}).
The evolution of $\Omega_{\rm DE}$ and $w_{\rm DE}$ is 
known accordingly.

Let us consider the interacting theories given by 
the function
\be
f=-f_1(\phi,X,Z) \rho_c(n_c)+f_2(\phi,X,Z)\,,
\label{fint}
\ee
where $f_1$ and $f_2$ depend on $\phi,X,Z$. 
In this case, the coupling $f_1$ gives rise to the terms 
$f_1 \rho_c$ and $f_1 P_c$ in $\rho_{\rm DE}$ and 
$P_{\rm DE}$, respectively.
If these terms are absorbed into $\rho_c$ and $P_c$ 
appearing in Eqs.~(\ref{back1}) and (\ref{back2}), respectively, 
then we can define the effective CDM density and pressure, as 
\be
\hat{\rho}_c=\left( 1+f_1 \right) \rho_c\,,\qquad
\hat{P}_c=\left( 1+f_1 \right) P_c\,,
\label{hatrhoc}
\ee
together with 
\be
\hat{\rho}_{\rm DE}=\rho_{\rm DE}-f_1\rho_c\,,\qquad
\hat{P}_{\rm DE}=P_{\rm DE}-f_1P_c\,.
\label{hatde}
\ee
Then, from the continuity Eqs.~(\ref{coneq}) and (\ref{coneq2}), 
it follows that 
\ba
& &
\dot{\hat{\rho}}_c+3H \left( \hat{\rho}_c+\hat{P}_c \right)
=+\frac{\dot{f}_1}{1+f_1} \hat{\rho}_c\,,
\label{hatrhoceq}\\
& &
\dot{\hat{\rho}}_{\rm DE}+3H \left( \hat{\rho}_{\rm DE}
+\hat{P}_{\rm DE} \right)
=-\frac{\dot{f}_1}{1+f_1} \hat{\rho}_c\,,
\label{hatPceq}
\ea
whose right-hand-sides are opposite to each other. 
Hence the energy exchange between CDM and DE
is explicit with the definitions (\ref{hatrhoc}) and (\ref{hatde}).
The CDM density $\rho_c$ and pressure $P_c$ are those 
associated with the conservation of CDM particle number 
$J_c^0={\cal N}_c$, 
so the standard continuity equation (\ref{coneq}) holds for them.  
The CDM acquires a field-dependent effective mass through  
the energy exchange with the scalar field. 
This results in the modified continuity Eq.~(\ref{hatrhoceq}).

\section{Second-order action and perturbation equations}
\label{S2sec}

In this section, we first derive the second-order action of tensor 
perturbations on the flat FLRW background for the coupled DE and DM theories 
given by the action (\ref{action}). 
Then, we proceed to the derivation of the second-order action of 
scalar perturbations. There are no dynamical vector degrees of freedom 
for the theories under consideration, so we do not take vector 
perturbations into account.

The matter sector is dealt as perfect fluids described by the 
Schutz-Sorkin action, so there are no anisotropic shear and viscosity.
The matter components interact with each other only through gravity 
except for the interaction between the scalar field and CDM. 
This perfect-fluid treatment for CDM and baryons is sufficient 
to study the late-time dynamics of perturbations associated 
with the cosmic growth rate.

\subsection{Tensor perturbations}
\label{Appen}

The perturbed line element containing the tensor perturbation 
$h_{ij}$ on the flat FLRW background is given by 
\be
{\rm d} s^2=-{\rm d} t^2+a^2(t) \left( \delta_{ij}
+h_{ij} \right) {\rm d}x^i {\rm d}x^j\,,
\ee
where $h_{ij}$ obeys the transverse and traceless conditions 
$\partial^i h_{ij}=0$ and $h^i_i=0$. 
We consider the gravitational waves propagating along the 
$z$ direction, in which case the nonvanishing components of $h_{ij}$ 
can be chosen as $h_{11}=h_1(t,z)$, $h_{22}=-h_1(t,z)$, and 
$h_{12}=h_{21}=h_2(t,z)$. 
Expanding the action (\ref{action}) up to second order 
in $h_1$ and $h_2$, integrating it by parts, and
using the background equations of motion,  
the second-order action of tensor perturbations 
reduces to 
\be
{\cal S}_t^{(2)}=\int {\rm d}t\,{\rm d}^3x 
\sum_{i=1}^{2}
\frac{a^3}{4}q_t \left[ \dot{h}_i^2-\frac{c_t^2}{a^2} 
(\partial h_i)^2 \right]\,,
\label{actionSt}
\ee
where 
\be
q_t=2G_4\,,
\qquad
{\rm and}\qquad
c_t^2=1\,.
\label{qtct}
\ee
The tensor ghost is absent under the condition $G_4>0$. 
Since the speed of gravitational waves is equivalent to that of light,  
the theories given by the action  (\ref{action}) are consistent 
with the observational bound of $c_t^2$ derived from 
the GW170817 event \cite{GW170817}.

\subsection{Scalar perturbations}

The line element containing four scalar perturbations 
$\alpha,\chi,\zeta$ and $E$ is given by \cite{Bardeen} 
\be
{\rm d}s^2=-(1+2\alpha) {\rm d}t^2
+2 \partial_i \chi {\rm d}t {\rm d}x^i
+a^2(t) \left[ (1+2\zeta) \delta_{ij}
+2\partial_i \partial_j E \right] {\rm d}x^i {\rm d}x^j\,,
\label{permet}
\ee
where the perturbed quantities depend on both  cosmic time 
$t$ and spatial coordinates $x^i$. 

The scalar field is decomposed into the time-dependent background part 
$\bar{\phi}(t)$ and the perturbed part $\delta\phi$, as
\be
\phi=\bar{\phi}(t)+\delta\phi\,,
\label{dphi}
\ee
where we will omit the bar in the following discussion.
We also decompose the temporal and spatial components 
of $J_I^{\mu}$ in the forms, 
\be
J_I^{0}={\cal N}_I+\delta J_I\,,\qquad 
J_I^{i}=\frac{1}{a^2(t)}\delta^{ik}\partial_k\delta j_I\,, 
\label{JI}
\ee
where ${\cal N}_I$ is the background conserved
particle number, while $\delta J_I$ and $\delta j_I$ are 
the scalar perturbations. 

Substituting Eq.~(\ref{JI}) into the definition of number density 
(\ref{ndef}) and expanding $n_I$ up to second order in scalar perturbations, 
it follows that 
\be
n_I=\frac{{\cal N}_I}{a^3} \left[
1+\frac{\delta\rho_I}{\rho_{I}+P_I}\left(1-3\zeta-\partial^2 E\right)
-\frac{(\partial \delta j_I+{\cal N}_I \partial\chi)^2}{2 {\cal N}_I^2 a^2}
-\frac{1}{2}(\zeta+\partial^2E)(3\zeta-\partial^2E)\right]
+{\cal O}(\varepsilon^3)
\,, \label{deln}
\ee
where $\varepsilon$ represents the order of perturbations, and 
$\delta\rho_I$ is the density perturbation defined by 
\be
\delta\rho_I=\frac{\rho_{I}+P_I}{{\cal N}_I} \left[ \delta J_I-{\cal N}_I 
\left( 3\zeta+\partial^2 E \right) \right]\,.
\label{drhoI}
\ee
At linear order, the perturbation $\delta n_I$ of number density 
is related to $\delta\rho_I$ according to 
$\delta\rho_I=\rho_{I,n_I}\delta n_I$. 
By using Eqs.~(\ref{JI}) and (\ref{deln}), the four velocity 
$u_{I \mu}=J_{I \mu}/(n_I \sqrt{-g})$, 
which is expanded up to linear order, is given by 
\be
u_{I0}=-1-\alpha\,,\qquad
u_{Ii}=-\partial_i v_I\,, 
\label{uI}
\ee
where 
\be
\partial_i v_I=-\partial_i \left(\chi+\frac{\delta j_I}{{\cal N}_I}\right)\,.
\label{vI}
\ee
Note that $v_I$ corresponds to the velocity potential 
of each fluid. 
In the following, we express $\delta j_I$ and $\delta J_I$ in terms of $\delta \rho_I$, 
$v_I$, and metric perturbations. 
On using Eqs.~(\ref{dphi}) and (\ref{uI}), the spatial component 
of Eq.~(\ref{ellc}) expanded up to linear order in perturbations is given by 
\be
\partial_i \ell_c = -(\rho_{c,n_c}-f_{,n_c})\partial_i v_c
+\frac{a^3f_{,Z}}{{\cal N}_c}(\partial_i\delta\phi-\dot{\phi}\partial_i v_c)\,, 
\label{ellc2}
\ee
where the coefficients $\rho_{c,n_c},f_{,n_c},$ and $f_{,Z}$ should be
evaluated on the background. Integrating Eq.~(\ref{ellc2}) with respect to 
$x^i$ and using the property $\dot{\ell}_c=-(\rho_{c,n_c}-f_{,n_c})$ 
on the background, we obtain
\be
\ell_c=-\int^t [\rho_{c,n_c}(\tilde{t})-f_{,n_c}(\tilde{t})] {\rm d}\tilde{t}
-(\rho_{c,n_c}-f_{,n_c}) v_c
+\frac{a^3f_{,Z}}{{\cal N}_c}(\delta\phi-\dot{\phi} v_c)\,.
\ee
This relation will be used to eliminate the Lagrange multiplier $\ell_c$
from the action (\ref{action}).

\subsubsection{Second-order action}

Since the energy density $\rho_I$ depends on $n_I$, 
it can be expanded in the form, 
\be
\rho_I (n_I)=\rho_I+\left( \rho_I+P_I \right) 
\frac{\delta n_I}{n_I}+\frac{1}{2} \left( \rho_I+P_I \right) 
c_I^2 \left( \frac{\delta n_I}{n_I} \right)^2
+{\cal O} (\varepsilon^3)\,,
\ee
where $c_I^2$ is the fluid sound speed squared 
defined by 
\be
c_I^2=\frac{n_I \rho_{I,n_I n_I}}{\rho_{I,n_I}}\,.
\ee
We also express the interacting Lagrangian $f(n_c,\phi,X,Z)$ as
\ba
\hspace{-0.5cm}
f(n_c,\phi,X,Z)&=&f
+f_{,n_c} \delta n_c+f_{,\phi} \delta \phi+f_{,X} \delta X+f_{,Z} \delta Z
+\frac{1}{2} f_{,n_cn_c} \delta n_c^2+\frac{1}{2} f_{,\phi\phi} \delta \phi^2
+\frac{1}{2} f_{,XX} \delta X^2+\frac{1}{2} f_{,ZZ} \delta Z^2\notag\\
\hspace{-0.5cm}
&&+f_{,n_c\phi} \delta n_c \delta \phi+f_{,n_cX} \delta n_c \delta X
+f_{,n_cZ} \delta n_c \delta Z+f_{,X\phi} \delta \phi \delta X
+f_{,Z\phi} \delta \phi \delta Z+f_{,XZ} \delta X \delta Z+{\cal O} (\varepsilon^3)\,,
\hspace{.5cm}
\ea
where $\delta n_c$ is the perturbed part of Eq.~(\ref{deln}) 
with $I=c$, and 
\ba
& &
\delta X=\dot{\phi} (\dot{\delta \phi}-\dot{\phi} \alpha)
+\frac{1}{2} \left[  (\dot{\delta \phi}-2\dot{\phi} \alpha)^2
-\frac{1}{a^2}(\partial \delta \phi+\dot{\phi} \partial \chi)^2
\right]+{\cal O} (\varepsilon^3)\,,
\label{deltaX}\\
& &
\delta Z=\dot{\delta \phi}-\dot{\phi} \alpha
+\frac{1}{2a^2} \left[ \dot{\phi} \left\{ 3a^2 \alpha^2
-(\partial_i \chi)^2+(\partial_i v_c)^2 \right\}-2a^2 \alpha \dot{\delta \phi}
-2\partial_i \delta \phi (\partial_i \chi+\partial_i v_c) 
\right] +{\cal O} (\varepsilon^3)\,.
\label{delZ}
\ea

We expand the action (\ref{action}) up to quadratic order in scalar perturbations, 
integrate it by parts, and use the background equations of motion. 
Then, the resulting second-order action is expressed in the form, 
\be
{\cal S}_s^{(2)}=\int {\rm d}t\,{\rm d}^3x \left( L_{0}+L_{f} \right)\,,
\label{faction}
\ee
where
\ba
L_0
&=& a^3\Biggl\{
D_1\dot{\dphi}^2+D_2\frac{(\partial\dphi)^2}{a^2}+D_3\dphi^2
+\left(D_4\dot{\dphi}+D_5\dphi+D_6\frac{\partial^2\dphi}{a^2}\right) \alpha
-\left(D_6\dot{\dphi}-D_7\dphi\right)\frac{\partial^2\chi}{a^2}
 \nonumber \\
&&
+\left(\dot{\phi} D_6-2Hq_t\right)\alpha\frac{\partial^2\chi}{a^2}
+\left(\dot{\phi}^2 D_1+3H \dot{\phi} D_6-3H^2 q_t\right)\alpha^2
 \nonumber \\
&&
+\sum_{I=c,b,r}\left\{ \left( \rho_I+P_I \right)v_I \frac{\partial^2 \chi}
{a^2}-v_I \dot{\delta \rho}_I-3H (1+c_I^2) v_I \delta \rho_I 
-\frac{\rho_I+P_I }{2a^2} (\partial v_I)^2
-\frac{c_I^2}{2 (\rho_I+P_I)} \delta \rho_I^2 
-\alpha \delta \rho_I \right\} 
\nonumber \\
&&
+\biggl\{
3D_6\dot{\dphi}-3D_7\dphi-3\left(\dot{\phi} D_6-2Hq_t\right)\alpha
-\sum_{I=c,b,r} 3(\rho_I+P_I)v_I
+2q_t\frac{\pa^2\chi}{a^2}\biggr\}\dot{\zeta}-3q_t\dot{\zeta}^2
+q_t \frac{(\pa\zeta)^2}{a^2}
\nonumber \\
&&
-2\left(\frac{\dot{q}_t}{\tp}\dphi+q_t\alpha\right)\frac{\pa^2\zeta}{a^2}
+\biggl[D_6\dot{\dphi}-2q_t\dot{\zeta}-D_7\dphi-(\tp D_6-2Hq_t)\alpha
-\sum_{I=c,b,r} (\rho_I+P_I)v_I
\biggr]\pa^2 \dot{E} \Biggr\}\,,
\label{L0}
\ea
and 
\ba
L_{f}
&=&a^3\left\{
\frac12(f_{,X}+\tp^2 f_{,XX}+2\tp f_{,XZ}+f_{,ZZ})
\left( \dot{\dphi}-\tp\alpha \right)^2
-\frac12\left[(f_{,X\phi}+\tp^2f_{,XX\phi}+2\tp f_{,XZ\phi}+f_{,ZZ\phi})\ttp
\right.\right.\notag\\
&&\left.\left.
+\tp^2f_{,X\phi\phi}+3H\tp(f_{,X\phi}-n_c f_{,n_cX\phi})+\tp f_{,Z\phi\phi}
+3H(f_{,Z\phi}-n_cf_{,n_cZ\phi})-f_{,\phi\phi}\right]\dphi^2
-\frac{f_{,X}}{2a^2}\left[(\partial\dphi)^2-2\tp\dphi\pa^2\chi\right]
\right.\notag\\
&&\left.
+(f_{,\phi}-\tp^2f_{,X\phi}-\tp f_{,Z\phi})\alpha\dphi
-\frac{n_cf_{n_c}-\tp f_{,Z}}{\rho_c+P_c}
\left[ (\rho_c+P_c )v_c \frac{\partial^2 \chi}
{a^2}-v_c \dot{\delta \rho}_c-3H (1+c_c^2) v_c \delta \rho_c 
-\frac{\rho_c+P_c }{2a^2} (\partial v_c)^2\right] 
\right.\notag\\
&&\left.
+\frac{n_c^2f_{,n_cn_c}}{2(\rho_c+P_c)^2}\drhoc^2
+\frac{n_c(f_{,n_c}-\tp^2f_{,n_cX}-\tp f_{,n_cZ})}{\rho_c+P_c}\alpha\drhoc
-\frac{f_{,Z}-n_c(f_{,n_cZ}+\tp f_{,n_c X})}{\rho_c+P_c}\dot{\dphi}\drhoc
\right.\notag\\
&&\left.
-\frac{1}{\rho_c+P_c}
\left[(\tp f_{,XZ}+f_{,ZZ})\ttp+\tp f_{,Z\phi}+3H(f_{,Z}-n_cf_{,n_cZ})
-n_cf_{,n_c\phi}\right]\drhoc\dphi
\right.\notag\\
&&\left.
+\left[(n_cf_{,n_c}-\tp f_{,Z})v_c-\tp f_{,X}\dphi\right]\left(3\dot{\zeta}+\pa^2\dot{E}\right)
\right\}\,,
\ea
where $n_c$ is evaluated on the background, i.e., $n_c={\cal N}_c/a^3$. 
The Lagrangian $L_f$ arises from the coupling $f$. 
The quantity $q_t$ is given in Eq.~(\ref{qtct}).
As we derived in Sec.~\ref{Appen}, we require the condition $q_t>0$ 
to avoid the tensor ghost.
The other coefficients $D_i$ $(i=1,...,7)$ are given by
\ba
D_1 &=& \frac{1}{2} G_{2,X}+G_{3,\phi}+\frac{1}{2} \dot{\phi}^2 
\left( G_{2,XX}+G_{3,X \phi} \right) 
-\frac{3}{2} H \dot{\phi} \left( 2 G_{3,X} +\dot{\phi}^2 
G_{3,XX} \right)\,,
\label{D1}\\
D_2 &=& -\frac12 G_{2,X}-G_{3,\phi}+2 H{\tp} G_{3,X}
+\frac12 {\tp}^{2}G_{3,X\phi}
+\frac12\left( 2G_{3,X}+G_{3,{XX}}{\tp}^{2} \right) {\ttp}
\,,\label{D2}\\
D_3 &=& \frac12 G_{{2,\phi\phi}} 
-\frac12  \left(  
G_{{2,X\phi\phi}}+  G_{{3,\phi\phi\phi}} \right) {{\tp}}^{2}
+ \frac32\left( G_{{3,X\phi\phi}}{{\tp}}^{2}- G_{{2,X
\phi}}-2  G_{{3,\phi\phi}}  \right) H {\tp} 
\notag\\
&&
+ \frac32\left( {{\tp}}^{2}G_{{3,X\phi}}+2
G_{{4,\phi\phi}} \right) {\dot{H}}
+ \frac32\left( 
3{{\tp}}^{2}G_{{3,X\phi}}+4  G_{{4,\phi\phi}} \right) {H}^{2}
\notag\\
&&
-\left[ \frac12G_{{2,X\phi}}+G_{{3,\phi\phi}} 
-\frac32\left( G_{{3,{XX}  \phi}}{{\tp}}^{2}
+2  G_{{3,X\phi}} \right) H {\tp}+\frac12 \left( 
G_{{2,{XX}  \phi}}+ G_{{3,X\phi\phi}} \right) {{\tp}}^{2}
 \right] {\ttp}\,,
\label{D3}\\
D_4 &=& - \left( G_{2,X}+2\,G
_{3,\phi} \right) {\tp}
-\left( G_{2,{
XX}}+G_{3,X\phi} \right) {\tp}^{3}
+3\left( 3G_{3,X}{\tp}^{2}+G_{3,{XX}}{\tp}^{4}
+2G_{4,\phi} \right) H\,,
\label{D4}\\
D_5 &=& G_{2,\phi}-\dot{\phi}^2 \left( G_{2,X \phi}+G_{3,\phi \phi} 
\right)+3H \dot{\phi} \left( \dot{\phi}^2 G_{3,X \phi}
+2G_{4, \phi \phi} \right)+6H^2 G_{4, \phi}\,,
\label{D5}\\
D_6 &=& -\dot{\phi}^2 G_{3,X}-2G_{4, \phi}\,,
\label{D6}\\
D_7 &=& \dot{\phi} \left( G_{2,X}+2G_{3,\phi}+2G_{4, \phi \phi} 
\right)-H \left( 3 \dot{\phi}^2 G_{3, X}+2G_{4, \phi} \right)\,.
\label{D7}
\ea
Among these coefficients, there are following four conditions: 
\ba
&&2\dot{\phi}^2 D_2 =
-2H \dot{q}_t -\dot{\phi} 
\left( \dot{D}_6+H D_6+D_7 \right)\,,\label{D2re} \\
&&D_4=-2\dot{\phi}D_1-3HD_6\,,
\label{D4re}\\
&&2q_t \dot{H}-D_6 \ddot{\phi}+f_{,X}\tp^2
+\left( D_7+f_{,Z} \right) \dot{\phi}+\sum_{I=c,b,r}(\rho_I+P_I)-n_cf_{,n_c}=0\,,
\label{bg11}\\
&&
\left( 2D_1+f_{,X}+\dot{\phi}^2f_{,XX}+2\dot{\phi}f_{,XZ}+f_{,ZZ} 
\right)\ddot{\phi}
+3D_6\dot{H}-D_5+3HD_7\notag\\
&&
-f_{,\phi}+\dot{\phi} \left[3H(f_{,X}-n_cf_{,n_cX})+f_{,Z\phi}
+\dot{\phi}f_{,X\phi} \right]
+3H \left( f_{,Z}-n_cf_{,n_cZ} \right)=0\,,
\label{bgphi}
\ea
where Eq.~(\ref{bg11}) is equivalent to the subtraction of 
Eq.~(\ref{back2}) from Eq.~(\ref{back1}), and Eq.~(\ref{bgphi}) 
corresponds to the scalar-field Eq.~(\ref{coneq2}). 
In Sec.~\ref{gaugeinasec}, we will use these relations for simplifying the 
perturbation equations of motion.

Among scalar perturbations in the second-order action (\ref{faction}),
the variables $\alpha,\chi,v_c,v_b,v_r$ and $E$ are non-dynamical. 
We introduce the comoving wavenumber $k$ and derive the perturbation 
equations for these non-dynamical variables in Fourier 
space. Variations of the action (\ref{faction}) with respect to 
$\alpha,\chi,v_c,v_b,v_r,E$ lead to 
\ba
\hspace{-0.5cm}
& & \left[ D_4-\dot{\phi} ( f_{,X}+\dot{\phi}^2 f_{,XX}+2\tp f_{,XZ}+f_{,ZZ} )\right] 
(\dot{\delta \phi}-\tp\alpha)-3 \left(\dot{\phi} D_6-2Hq_t \right)
(\dot{\zeta}-H\alpha)
+\left( D_5 + f_{,\phi}-\dot{\phi}^2 f_{,X \phi} - \dot{\phi} f_{,Z \phi} \right) \dphi 
\notag\\
\hspace{-0.5cm}
&&
+\frac{k^2}{a^2}\left[ 2q_t\zeta-\left(\dot{\phi} D_6-2Hq_t\right)\left(\chi-a^2\dot{E}\right)
-D_6\dphi \right]-\sum_{I=c,b,r} \delta \rho_I 
+\frac{n_c(f_{,n_c}-\tp^2 f_{,n_cX}-\tp f_{,n_cZ})}{\rho_c+P_c}
\delta \rho_c=0\,,\label{pereq1} \\
\hspace{-0.5cm}
& &
D_6\dot{\dphi}-2q_t\dot{\zeta}
-\left(D_7+\dot{\phi}f_{,X} \right) \dphi-\left(\dot{\phi} D_6-2Hq_t\right)\alpha
-\sum_{I=c,b,r} \left( \rho_I+P_I \right)v_I
+\left(n_cf_{,n_c} -\tp\,f_{,Z} \right) v_c=0\,,
\label{eqchi}\\
\hspace{-0.5cm}
& &
\dot{\delta \rho}_I+3H \left( 1+c_I^2 \right) \delta \rho_I
+3 \left (\rho_I+P_I \right) \dot{\zeta}
+\frac{k^2}{a^2} \left( \rho_I+P_I \right) 
\left( v_I+\chi-a^2\dot{E} \right)=0\,, 
\qquad {\rm for} \quad I=c,b,r\,,\label{pereq3}\\
\hspace{-0.5cm}
& &
\dot{{\cal W}}+3H{\cal W}=0\,,
\label{eqE}
\ea
respectively, where 
\be
{\cal W}=2q_t\dot{\zeta}-D_6\dot{\dphi}+\left(D_7+\tp f_{,X} \right)\dphi
+\left(\tp D_6-2Hq_t \right)\alpha
+\sum_{I=c,b,r} (\rho_I+P_I)v_I-\left(n_cf_{,n_c}-\tp f_{,Z}\right)v_c\,.
\label{Wdef}
\ee
Varying the action (\ref{faction}) with respect to the dynamical 
perturbations $\dphi, \delta\rho_c, \delta\rho_b, \delta\rho_r, \zeta$, 
we obtain 
\ba
& &
\dot{\cal Z}+3H {\cal Z}+3 \left(D_7+\dot{\phi} f_{,X} \right)\dot{\zeta}
+M_{\phi}^2 \delta \phi 
-\left(D_5+f_{,\phi}-\tp^2f_{,X\phi}-\tp f_{,Z\phi} \right) \alpha 
\nonumber \\
& &
+\frac{1}{\rho_c+P_c}\left[\left( \tp f_{,XZ}+f_{,ZZ} \right)\ttp+\tp f_{,Z\phi}
+3H \left( f_{,Z}-n_cf_{,n_cZ} \right)-n_cf_{,n_c\phi}\right]\drhoc
\notag\\
 &&
-\frac{k^2}{a^2} \left[ 2D_2 \delta \phi -D_6 \alpha 
-D_7 \chi+\frac{2\dot{q}_t}{\dot{\phi}}\zeta-a^2\left(\dot{D}_7+3HD_7 \right)E
-f_{,X}\left\{ \delta \phi+\dot{\phi} \left(\chi-a^2\dot{E} \right)  
\right\} \right]=0\,,\label{calZeq}\\
&&
\left(1+\frac{\tp f_{,Z}-n_cf_{,n_c}}{\rho_c+P_c}\right)\dot{v}_c
-\left(c_c^2-\frac{n_c^2f_{,n_cn_c}}{\rho_c+P_c}\right)
\left(3Hv_c+\frac{\drhoc}{\rho_c+P_c}\right)
\notag\\
&&
-\frac{1}{\rho_c+P_c}\left[ \left(\tp f_{,XZ}+f_{,ZZ} \right)\ttp+\tp f_{,Z\phi}
+3H\left(f_{,Z}-n_cf_{,n_cZ} \right)-n_cf_{,n_c\phi}\right](\dphi-\tp v_c)
\notag\\
&&
-\frac{f_{,Z}-n_c(\tp f_{,n_cX}+f_{,n_cZ})}{\rho_c+P_c}(\dot{\dphi}-\ttp v_c)
-\left[1-\frac{n_c(f_{,n_c}-\tp^2f_{,n_cX}-\tp f_{,n_cZ})}{\rho_c+P_c}\right]\alpha
=0\,,
\label{vceq}\\
\hspace{-1cm}
& &
\dot{v}_I-3H c_I^2\,v_I-\frac{c_I^2}{\rho_I+P_I} \delta \rho_I 
-\alpha=0\,,\qquad \qquad {\rm for} \quad I=b,r\,,
\label{veq}\\
\hspace{-1cm}
& &
\dot{{\cal W}}+3H{\cal W}
+\frac{2k^2}{3a^2}\left\{q_t \left[ \alpha+\dot{\chi}
+\zeta+H\chi-a^2 \left( \ddot{E}+3 H \dot{E} \right)  
\right]+  \dot{q}_t \left( \chi-a^2 \dot{E} 
+\frac{\dphi}{\dot{\phi}} \right)\right\}=0\,,
\label{calWeq}
\ea
where 
\ba
M_{\phi}^2 &=& 
-2D_3-f_{,\phi\phi}+\left( f_{,X\phi}+\tp^2f_{,XX\phi}+2\tp f_{,XZ\phi}
+f_{,ZZ\phi} \right)\ttp
+\tp^2f_{,X\phi\phi}+\tp f_{,Z\phi\phi}
\notag\\
&&
+3H\tp \left(f_{,X\phi}-n_c f_{,n_cX\phi} \right)
+3H \left(f_{,Z\phi}-n_cf_{,n_cZ\phi} \right)\,,\\
{\cal Z} &=& 2D_1\dot{\delta \phi}+\left(f_{,X}+\dot{\phi}^2 f_{,XX}
+2\tp f_{,XZ}+f_{,ZZ} \right) (\dot{\delta \phi}-\tp\alpha)+3D_6\dot{\zeta}
+D_4\alpha \notag\\
&&
-\frac{f_{,Z}-n_c(\tp f_{,n_cX}+f_{,n_cZ})}{\rho_c+P_c}\delta \rho_c 
+\frac{k^2}{a^2}\left[D_6\chi 
-a^2 \left(D_6\dot{E}+D_7E \right)\right]\,. 
\label{Zdef}
\ea
The quantity $M_{\phi}^2$ corresponds to the effective mass squared 
of scalar field perturbation. 
On using Eqs.~(\ref{eqE}) and (\ref{calWeq}), it follows that  
\be
q_t \left[ \alpha+\dot{\chi}
+\zeta+H\chi-a^2 \left( \ddot{E}+3 H \dot{E} \right)  
\right]+  \dot{q}_t \left( \chi-a^2 \dot{E} 
+\frac{\dphi}{\dot{\phi}} \right)=0\,. 
\label{eqE2}
\ee
Since we have not yet fixed the gauge degrees of freedom, 
the perturbation equations (\ref{pereq1})-(\ref{calWeq}) can be applied 
to any choices of gauges.  Namely, they are written in a gauge-ready 
form \cite{Hwang:2001qk,Heisenberg:2018wye}. 

\subsubsection{Perturbation equations with gauge-invariant variables}
\label{gaugeinasec}

In this subsection, we rewrite the perturbation equations of motion in terms of 
gauge-invariant variables. Let us consider the infinitesimal 
transformation given by 
\be
\tilde{t}=t+\xi^{0}\qquad 
{\rm and}\qquad 
\tilde{x}^{i}=x^{i}+\delta^{ij} \partial_{j} \xi\,, 
\label{trans}
\ee
where $\xi^{0}$ and $\xi$ are scalar variables. 
Then, the metric perturbations in Eq.~(\ref{permet}) 
transform as  
\be
\tilde{\alpha}=\alpha-\dot{\xi}^{0}\,,\qquad 
\tilde{\chi}=\chi+\xi^{0}-a^2 \dot{\xi}\,,\qquad 
\tilde{\zeta}=\zeta-H \xi^{0}\,,\qquad 
\tilde{E}=E-\xi\,,
\label{gaugetra1}
\ee
while the perturbations associated with the scalar field and 
fluids transform as
\be
\widetilde{\delta \phi}=\delta \phi-\dot{\phi}\,\xi^{0}\,,
\qquad 
\widetilde{\delta \rho_I}=\delta \rho_I-\dot{\rho}_I \xi^{0}\,,
\qquad 
\tilde{v}_I=v_I-\xi^{0}\,.
\label{gaugetra2}
\ee
We introduce the following variables invariant under the transformation 
(\ref{trans}), 
\ba
& &
\Psi=\alpha+\frac{{\rm d}}{{\rm d}t} 
\left( \chi - a^2 \dot{E} \right)\,,\qquad 
\Phi=\zeta+H \left( \chi - a^2 \dot{E} \right)\,,
\nonumber \\
& &
\delta \phi_{\rm N}=\delta \phi+\dot{\phi}\left(\chi-a^2 \dot{E}\right)\,,
\qquad 
\delta\rho_{I\rm N}=\delta \rho_I+\dot{\rho}_I \left(\chi-a^2 \dot{E}\right)\,,
\qquad 
v_{I{\rm N}}=v_I+\chi-a^2 \dot{E}\,,
\label{delphiN}
\ea
where $\Psi$ and $\Phi$ are Bardeen gravitational potentials \cite{Bardeen}. 
To simplify the perturbation equations of motion, we also define
the dimensionless variables,
\ba
&&
\aK=\frac{2\tp^2 D_1}{H^2q_t}\,,\qquad
\aB=-\frac{\tp D_6}{2Hq_t}\,,\qquad
\aM=\frac{\dot{q_t}}{Hq_t}\,,\notag\\ 
&&
\bK=\frac{\tp^2(f_{,X}+\tp^2f_{,XX}+2\tp f_{,XZ}+f_{,ZZ})}{H^2q_t}\,,\qquad 
\bn=\frac{n_c(f_{,n_c}-\tp^2f_{,n_cX}-\tp f_{,n_cZ})}{\rho_c+P_c}\,,
\label{nodim}
\ea
and 
\be
\epsilon_{\aK}=\frac{\dot{\aK}}{H\aK}\,,\qquad
\epsilon_{\aB}=\frac{\dot{\aB}}{H{\aB}}\,,\qquad
\epsilon_{\bK}=\frac{\dot{\bK}}{H\bK}\,,\qquad
\epsilon_{\beta_{n_c}}=\frac{\dot{\beta}_{n_c}}{H \beta_{n_c}}\,,\qquad
\eH=\frac{\dot{H}}{H^2}\,,\qquad 
\ep=\frac{\ddot{\phi}}{H\tp}\,.
\ee
The quantity $\aB$ is related to $\aB^{\rm (BS)}$ introduced 
by Bellini and Sawicki \cite{Bellini}, 
as $\aB=-\aB^{\rm (BS)}/2$, while $\aK$ and $\aM$ are 
the same as those given in Ref.~\cite{Bellini}.
The quantities $\bK$ and $\bn$ are 
new dimensionless variables arising from the coupling $f$.

In the following, we eliminate $D_2,D_4,D_5,D_7$ by using the relations 
(\ref{D2re})-(\ref{bgphi}), and replace $D_1,D_3,D_6$ with 
$\aK,M_{\phi}^2,\aB$, respectively.
On using the gauge-invariant variables given in Eq.~(\ref{delphiN}), 
the equations of motion for the non-dynamical perturbations 
$\alpha,\chi,v_c, v_b, v_r, E$, i.e., Eqs.~(\ref{pereq1})-(\ref{eqE}), 
and for the dynamical perturbations 
$\dphi, \delta\rho_c, \delta\rho_b, \delta\rho_r$, i.e., 
Eqs.~(\ref{calZeq})-(\ref{veq}), are expressed as
\ba
\hspace{-0.8cm}
&&
6(1+\aB)\frac{\dot{\Phi}}{H}+(6\aB-\aK-\bK)\frac{\dot{\dphi}_{\rm N}}{\tp}
+2 \left( \frac{k}{aH} \right)^2 \Phi-(6+12\aB-\aK-\bK)\Psi
-(1-\bn)\frac{\delta\rho_{c{\rm N}}}{H^2q_t}
-\sum_{I=b,r}\frac{\delta\rho_{I{\rm N}}}{H^2q_t}
\notag\\
\hspace{-0.8cm}
&&
+\left[2 \left( \frac{k}{aH} \right)^2\aB-6(1+\aB)\eH-(6\aB-\aK-\bK)\ep
-\frac{3(\rho_c+P_c)}{H^2q_t}(1-\bn)
-\sum_{I=b,r}\frac{3(\rho_I+P_I)}{H^2q_t}\right]\frac{H}{\tp}\dphi_{\rm N}=0\,,
\label{perteq1}\quad\\
\hspace{-0.8cm}
&&
\frac{\dot{\Phi}}{H}+\aB\frac{\dot{\dphi}_{\rm N}}{\tp}-(1+\aB)\Psi
+\frac{q_c(\rho_c+P_c)}{2Hq_t}\left(v_{c{\rm N}}-\frac{\dphi_{\rm N}}{\tp}\right)
+\sum_{I=b,r}\frac{\rho_I+P_I}{2Hq_t}\left(v_{I{\rm N}}-\frac{\dphi_{\rm N}}{\tp}\right)
\notag\\
\hspace{-0.8cm}
&&
-(\eH+\ep\aB)\frac{H}{\tp}\dphi_{\rm N}=0\,,\\
\hspace{-0.8cm}
&&
\dot{\delta\rho}_{I{\rm N}}+3H(1+c_I^2)\delta\rho_{I{\rm N}}
+(\rho_I+P_I)\left(3\dot{\Phi}+\frac{k^2}{a^2}v_{I{\rm N}}\right)=0\,, 
\qquad \quad {\rm for} \quad I=c,b,r, \label{delrhocN} \\
\hspace{-0.8cm}
&&
\dot{\cal W}+3H {\cal W}=0\,,
\label{cWeq}
\ea
and
\ba
\hspace{-1.3cm}
&&
(\aK+\bK)\frac{\ddot{\dphi}_{\rm N}}{H\tp}
-6\aB\frac{\ddot{\Phi}}{H^2}
+\left[\epsilon_{\aK}\aK+\epsilon_{\bK}\bK
+(3+\aM+2\eH-2\ep)(\aK+\bK)\right]\frac{\dot{\dphi}_{\rm N}}{\tp}
+(6\aB-\aK-\bK)\frac{\dot{\Psi}}{H}
\notag\\
\hspace{-1.3cm}
&&
-3\left[2\eH+2(3+\eH+\aM+\epsilon_{\aB})\aB
+\frac{q_c(\rho_c+P_c)}{H^2q_t}
+\sum_{I=b,r}\frac{\rho_I+P_I}{H^2q_t}\right]\frac{\dot{\Phi}}{H}
-2\aM \left( \frac{k}{aH} \right)^2 \Phi \notag\\
\hspace{-1.3cm}
&&
-\bigg[2\aB \left( \frac{k}{aH} \right)^2+\epsilon_{\aK}\aK
+\epsilon_{\bK}\bK-6\epsilon_{\aB}\aB
+(3+2\eH+\aM)(\aK+\bK-6\aB)-6\eH(1+\aB)
\notag\\
\hspace{-1.3cm}
&&
-3(1-\bn)\frac{\rho_c+P_c}{H^2q_t}-3\sum_{I=b,r}\frac{\rho_c+P_c}{H^2q_t}
\bigg]\Psi+\left[\left( \frac{k}{aH} \right)^2 \left\{2\aB(\aB-2\aM)
+\cfrac{\tp^2q_s\hat{c}_s^2}{2H^2q_t^2}\right\}
+\frac{\tp^2M_{\phi}^2}{H^4q_t}
\right]\frac{H}{\tp}\dphi_{\rm N} \notag\\
\hspace{-1.3cm}
&&
+(1-\beta_{n_c}-q_c )\frac{\dot{\delta \rho}_{c{\rm N}}}{H^3 q_t}
+\left[ 3-3q_c(1-\hat{c}_c^2+c_c^2)-(3+\epsilon_{\beta_{n_c}}) \beta_{n_c} \right]
\frac{\delta \rho_{c{\rm N}}}{H^2 q_t}=0\,,\label{delphieq}\\
\hspace{-1.3cm}
&&
\dot{v}_{c{\rm N}}-H(3c_c^2-\epsilon_{q_c})v_{c{\rm N}}
-\frac{\hat{c}_c^2\delta\rho_{c{\rm N}}}{\rho_c+P_c}
-\frac{(1-\bn)\Psi}{q_c}+\frac{1-\bn-q_c}{q_c \tp}
(\dot{\dphi}_{\rm N}-H\ep\dphi_{\rm N})
-(3\hat{c}_c^2-3c_c^2+\epsilon_{q_c})\frac{H\dphi_{\rm N}}{\tp}=0\,,
\label{vcNeq}\\
\hspace{-1.3cm}
&&
\dot{v}_{I{\rm N}}-3Hc_I^2v_{I{\rm N}}
-\frac{c_I^2}{\rho_I+P_I}\delta\rho_{I{\rm N}}-\Psi=0\,,
\qquad \quad {\rm for} \quad I=b,r,
\ea
where  
\ba
&&
q_c=1+\frac{\tp f_{,Z}-n_cf_{,n_c}}{\rho_c+P_c}\,,
\label{qc}\\
&&
q_s=\frac{2H^2q_t^2(\aK+\bK+6\aB^2)}{\tp^2}\,,
\label{qs}\\
&&
\hat{c}_c^2=\frac{1}{q_c}\left(c_c^2
-\frac{n_c^2f_{,n_cn_c}}{\rho_c+P_c}\right)\,, 
\label{hatcs2}\\ 
&&
\hat{c}_s^2=-\frac{4H^2q_t^2}{\tp^2q_s}\left[
\eH-\aM+\aB(1+\eH-\aM+\aB+\epsilon_{\aB})
+\frac{q_c(\rho_c+P_c)}{2H^2q_t}
+\sum_{I=b,r}\frac{\rho_I+P_I}{2H^2q_t}
\right]\,,
\label{hatcs}\\ 
&&
\epsilon_{q_c}=\frac{\dot{q}_c}{H q_c}\,,
\ea
and
\be
\frac{{\cal W}}{2Hq_t}=\aB\frac{\dot{\dphi}_{\rm N}}{\tp}+\frac{\dot{\Phi}}{H}
-(\eH+\ep\aB)\frac{H}{\tp}\dphi_{\rm N}
-(1+\aB)\Psi+\sum_{I=b,r}\frac{\rho_I+P_I}{2Hq_t}
\left(v_{I{\rm N}}-\frac{\dphi_{\rm N}}{\tp}\right)
+\frac{q_c(\rho_c+P_c)}{2Hq_t}\left(v_{c{\rm N}}
-\frac{\dphi_{\rm N}}{\tp}\right).
\label{Wq}
\ee
As we will see later in Sec.~\ref{stasec}, the quantities $q_c$, $q_s$, 
$\hat{c}_c^2$, and $\hat{c}_s^2$ are related to the stability conditions
of CDM and scalar-field perturbations. 
We note that Eq.~(\ref{cWeq}) is written in terms of the gauge-invariant 
variable ${\cal W}$ given by Eq.~(\ref{Wq}). 
Finally, Eq.~(\ref{eqE2}) is expressed as
\be
\Psi+\Phi+\aM\frac{H}{\tp}\dphi_{\rm N}=0\,. 
\label{aniso}
\ee
The equation of motion for $\zeta$, which is given by Eq.~(\ref{calWeq}), 
is the combination of Eqs.~(\ref{cWeq}) and (\ref{aniso}).  
{}From Eq.~(\ref{aniso}), it follows that there is an anisotropic stress 
($\Psi \neq -\Phi$) for the theories with $\aM \neq 0$.

\section{Stability conditions}
\label{stasec}

In this section, we derive the stability conditions for scalar perturbations deep inside the sound horizon. Since these conditions are independent of the choice of gauges \cite{Kase:2018aps}, 
the residual gauge degrees of freedom can be fixed by choosing a particular gauge.
Let us choose the unitary gauge characterized by 
\be
\dphi=0\,,\qquad E=0\,, 
\ee
which is realized by setting $\xi=E$ and $\xi^0=\dphi/\tp$ 
in Eqs.~(\ref{gaugetra1}) and (\ref{gaugetra2}), respectively. 
We also introduce the following gauge-invariant variables,
\be
{\cal R}=\zeta-\frac{H}{\dot{\phi}} \delta \phi\,,\qquad 
\delta \rho_{I{\rm u}}=\delta \rho_I -\frac{\dot{\rho}_I}
{\dot{\phi}}\delta \phi\,,
\ee
which reduce to ${\cal R}=\zeta$ and $\delta \rho_{I{\rm u}}=\delta\rho_I$ 
in the unitary gauge.

We solve Eqs.~(\ref{pereq1})-(\ref{pereq3}) for $\alpha$, $\chi$, 
$v_c$, $v_b$, $v_r$ to eliminate the non-dynamical variables 
from the second-order action (\ref{faction}).
After the integration by parts, the resulting second-order action 
for dynamical perturbations 
${\cal R}, \delta \rho_{c{\rm u}}, \delta \rho_{b{\rm u}}, 
\delta \rho_{r{\rm u}}$ is expressed in the form, 
\be
{\cal S}_s^{(2)}=\int {\rm d}t\,{\rm d}^3x\,a^{3}\left( 
\dot{\vec{\mathcal{X}}}^{t}{\bm K}\dot{\vec{\mathcal{X}}}
-\frac{k^2}{a^2}\vec{\mathcal{X}}^{t}{\bm G}\vec{\mathcal{X}}
-\vec{\mathcal{X}}^{t}{\bm M}\vec{\mathcal{X}}
-\frac{k}{a}\vec{\mathcal{X}}^{t}{\bm B}\dot{\vec{\mathcal{X}}}
\right)\,,
\label{Ss2}
\ee
where ${\bm K}$, ${\bm G}$, ${\bm M}$, ${\bm B}$ 
are $4 \times 4$ matrices, and
\be
\vec{\mathcal{X}}^{t}=\left( 
{\cal R},  
\delta \rho_{c{\rm u}}/k, 
\delta \rho_{b{\rm u}}/k, 
\delta \rho_{r{\rm u}}/k \right) \,.
\label{calX}
\ee
Taking the small-scale limit, the leading-order matrix components 
for ${\bm K}$, ${\bm G}$, ${\bm B}$ are given, respectively, by 
\ba
& &
K_{11}=\frac{q_s \dot{\phi}^2}{4H^2q_t(1+\aB)^2}\,,\qquad 
K_{22}=\frac{q_ca^2}{2(\rho_c+P_c)}
\,,\qquad 
K_{33}=\frac{a^2}{2(\rho_b+P_b)}\,,\qquad 
K_{44}=\frac{a^2}{2(\rho_r+P_r)}\,,
\label{Kcomp}
\\
& &
G_{11}=\frac{q_s\hat{c}_s^2\dot{\phi}^2}{4H^2q_t(1+\aB)^2}\,,\qquad 
G_{22}=\frac{\hat{c}_c^2 q_c\,a^2}{2(\rho_c+P_c)}\,,\qquad 
G_{33}=\frac{c_b^2\,a^2}{2(\rho_b+P_b)}\,,\qquad 
G_{44}=\frac{c_r^2\,a^2}{2(\rho_r+P_r)}\,,
\label{Gcomp}
\\
&&
B_{12}=-B_{21}=-\frac{a(1-\bn-q_c)}{2H (1+\aB)}\,,
\label{Bcomp}
\ea
where $q_c$, $q_s$, $\hat{c}_c^2$, and $\hat{c}_s^2$ are 
defined in Eqs.~(\ref{qc})-(\ref{hatcs}). 

To avoid the scalar ghosts, the components of ${\bm K}$ in 
Eq.~(\ref{Kcomp}) must be positive. 
As long as the ghost is absent in the tensor sector ($q_t>0$) 
and the weak energy conditions $\rho_I+P_I>0$ hold
for $I=c,b,r$, the no-ghost conditions are given by 
\be
q_s>0\qquad {\rm and} \qquad
q_c>0\,.
\ee

The dispersion relations for baryons and radiation are not affected by 
the off-diagonal components of matrix ${\bm B}$, so their propagation 
speed squares are given, respectively, by 
$c_b^2=G_{33}/K_{33}$ and $c_r^2=G_{44}/K_{44}$. 
The off-diagonal components (\ref{Bcomp}) can modify the propagation 
of perturbations ${\cal X}_1 \equiv {\cal R}$ and 
${\cal X}_2 \equiv \delta \rho_{c{\rm u}}/k$. 
We vary the second-order action (\ref{Ss2}) with respect to 
the variables ${\cal X}_j$ (where $j=1,2$) and then substitute 
the solutions of the form 
${\cal X}_j=\tilde{{\cal X}}_j e^{i (\omega t-kx)}$ 
into their equations of motion. 
In the small-scale limit, the dominant contributions to the 
dispersion relation are those containing 
$\omega^2$, $\omega k$, and $k^2$. 
Then, it follows that
\ba
& & \omega^2 \tilde{{\cal X}}_1-\hat{c}_s^2\frac{k^2}{a^2}
\tilde{{\cal X}}_1-i \omega \frac{k}{a} \frac{B_{12}}{K_{11}} 
\tilde{{\cal X}}_2  \simeq 0\,,\label{dis1}\\
& & \omega^2 \tilde{{\cal X}}_2-\hat{c}_c^2\frac{k^2}{a^2}
\tilde{{\cal X}}_2-i \omega \frac{k}{a} \frac{B_{21}}{K_{22}} 
\tilde{{\cal X}}_1 \simeq 0\,.
\label{dis2}
\ea

We will focus on the case in which the 
bare CDM sound speed squared vanishes, i.e., 
\be
c_c^2=\frac{n_c \rho_{c,n_c n_c}}{\rho_{c,n_c}} \to 0\,.
\ee
For the coupling $f$ obeying the condition, 
\be
f_{,n_cn_c}=0\,,
\label{fncnc}
\ee
we have $\hat{c}_c^2=0$ from Eq.~(\ref{hatcs2}).
In this case, we obtain the two separable solutions 
to Eq.~(\ref{dis2}), as
\ba
& &
\omega=0\,,\label{branch1}\\
& &
\omega \tilde{\cal X}_2-i \frac{k}{a}
\frac{B_{21}}{K_{22}}
\tilde{{\cal X}}_1=0\,.\label{branch2}
\ea
The dispersion relation (\ref{branch1}) is that of CDM, 
so the resulting CDM effective sound speed squared is given by 
\be
c_{\rm CDM}^2=\omega^2 \frac{a^2}{k^2}=0\,.
\ee

Substituting the other solution (\ref{branch2}) to Eq.~(\ref{dis1}), 
the dispersion relation for the perturbation ${\cal R}$ 
is expressed in the form $\omega^2=c_s^2k^2/a^2$, where
\be
c_s^2=\hat{c}_s^2+\Delta c_s^2\,,
\ee
with 
\be
\Delta c_s^2=\frac{B_{12}^2}{K_{11}K_{22}}
=\frac{2q_t(\rho_c+P_c)(1-\bn-q_c)^2}{q_s q_c\tp^2}\,.
\ee
We recall that $\hat{c}_s^2$ is given by Eq.~(\ref{hatcs}). 
The off-diagonal components of ${\bm B}$ give rise to 
the modification $\Delta c_s^2$ to $\hat{c}_s^2$. 
The Laplacian instability of the perturbation ${\cal R}$ is 
absent for
\be
c_s^2 \geq 0\,.
\label{cscon}
\ee
Under the absence of scalar and tensor ghosts, 
it follows that $\Delta c_s^2 \geq 0$. 
This means that even the negative value of $\hat{c}_s^2$ in the range 
$\hat{c}_s^2 \geq -\Delta c_s^2$ can satisfy the condition (\ref{cscon}). 
However, in the regime where the scalar field dominates over CDM, 
the term $\Delta c_s^2$ can be negligible relative to $\hat{c}_s^2$, 
so it is necessary to satisfy the condition $\hat{c}_s^2 \geq 0$ to ensure 
the stability during the whole cosmic expansion history. 

The theories obeying the condition (\ref{fncnc}) 
corresponds to the coupling $f$ containing the 
linear dependence of $n_c$, i.e., 
\be
f=-\tilde{f}_1(\phi,X,Z) n_c+f_2(\phi,X,Z)\,,
\label{fcon}
\ee
where $\tilde{f}_1(\phi,X,Z)$ and $f_2(\phi,X,Z)$ are arbitrary 
functions of $\phi$, $X$, and $Z$.
The CDM with mass $m_c$ has the density $\rho_c=m_c n_c$, so 
the coupling (\ref{fcon}) is equivalent to 
\be
f=-f_1(\phi,X,Z) \rho_c+f_2(\phi,X,Z)\,,
\label{fcon2}
\ee
where $f_1=\tilde{f}_1/m_c$. 
Since $c_{\rm CDM}^2=0$ in this case, there is no additional pressure which prevents 
or enhances the gravitational clustering of CDM density perturbations. 
In Refs.~\cite{Pourtsidou:2013nha,Boehmer:2015sha,Koivisto:2015qua,Pourtsidou:2016ico,Linton,Kase:2019veo,Kase:2019mox,Chamings:2019kcl,Amendola:2020ldb}, 
the authors studied the cosmology for several sub-classes of couplings 
which belong to the general form (\ref{fcon2}).

\section{Effective gravitational couplings}
\label{Geffsec}

We derive the effective gravitational couplings of CDM and baryons 
for the interacting theories satisfying  
\be
f_{,n_c n_c}=0\,,
\label{fnn}
\ee
together with the conditions, 
\be
P_c=0\,,\qquad c_c^2=0\,,\qquad
P_b=0\,,\qquad c_b^2=0\,.
\ee
In this case, the CDM effective sound speed squared 
$c_{\rm CDM}^2$ vanishes.
We also neglect the contribution of radiation to the dynamics of 
both background and perturbations.

To study the evolution of CDM and baryon density perturbations, we introduce
the gauge-invariant matter density contrast, 
\be
\delta_{I{\rm N}}=\frac{\delta \rho_{I\rm N}}{\rho_I}\,,
\ee
where $I=c,b$. 
{}From Eqs.~(\ref{delrhocN}) and (\ref{vcNeq}), 
the CDM density contrast $\delta_{c{\rm N}}$ and 
velocity potential $v_{c{\rm N}}$ obey
\ba
& &
\dot{\delta}_{c{\rm N}}+3\dot{\Phi}+\frac{k^2}{a^2}v_{c{\rm N}}=0\,,
\label{delceq}\\
& &
\dot{v}_{c{\rm N}}+H \epsilon_{q_c} v_{c{\rm N}}-\frac{1-\beta_{n_c}}{q_c}\Psi
+\frac{1-\beta_{n_c}-q_c}{q_c \dot{\phi}} \dot{\delta \phi}_{\rm N}
-\frac{H}{q_c \dot{\phi}} \left[ (1-\beta_{n_c}-q_c ) \epsilon_{\phi}
+q_c \epsilon_{q_c} \right] \delta \phi_{\rm N}=0\,.
\label{vceq2}
\ea
Differentiating Eq.~(\ref{delceq}) with respect to $t$ and using Eq.~(\ref{vceq2}), 
it follows that 
\ba
& &
\ddot{\delta}_{c{\rm N}}+\left( 2+\epsilon_{q_c} \right) H \dot{\delta}_{c{\rm N}}
+\frac{k^2}{a^2} \frac{1-\beta_{n_c}}{q_c}\Psi
-\frac{k^2}{a^2} \frac{1-\beta_{n_c}-q_c}{q_c} 
\frac{\dot{\delta \phi}_{\rm N}}{\dot{\phi}}
+\frac{k^2}{a^2} \frac{(1-\beta_{n_c}-q_c ) \epsilon_{\phi}
+q_c \epsilon_{q_c}}{q_c} \frac{H}{\dot{\phi}}\delta \phi_{\rm N} \nonumber \\
& &=
-3\ddot{\Phi}-3\left( 2+\epsilon_{q_c} \right) H\dot{\Phi}\,.
\label{quasi0}
\ea

Let us employ the quasi-static approximation for the perturbations 
deep inside the sound horizon, under which the dominant contributions to the 
perturbation equations are those containing $k^2$, $\delta_{c{\rm N}}$, 
$\dot{\delta}_{c{\rm N}}$, and 
$\delta_{b{\rm N}}$ \cite{Boisseau:2000pr,Tsujikawa:2007gd,DeFelice:2011hq}. 
We ignore the mass squared $M_{\phi}^2$ of the scalar degree of freedom 
arising from a scalar potential $V(\phi)$. 
This is a good approximation to study the evolution of perturbations 
in the late Universe, apart from dark energy models in which $M_{\phi}$ 
is much larger than $H$ until recently.
For later convenience, we introduce the following combinations, 
\be
\Delta_1=\alpha_{\rm B}-\alpha_{\rm M}\,,\qquad 
\Delta_2=\frac{\dot{\phi}^2 q_s \hat{c}_s^2}{4H^2 q_t^2} \,,\qquad
\Delta_3=(1-\beta_{n_c})\Delta_1-\beta_{n_c}\epsilon_{\beta_{n_c}} \,.
\label{Delta} 
\ee
Applying the quasi-static approximation to Eqs.~(\ref{perteq1}) 
and (\ref{delphieq}), it follows that 
\ba
\hspace{-0.5cm}
& &
2q_t \frac{k^2}{a^2} \left( \Phi+\frac{H \alpha_{\rm B}}{\dot{\phi}} 
\delta \phi_{\rm N} \right)-(1-\beta_{n_c}) \rho_c \delta_{c{\rm N}}
-\rho_b \delta_{b{\rm N}}=0\,,\\
\hspace{-0.5cm}
& &
2H q_t \frac{k^2}{a^2} \left[ (\Delta_1-\alpha_{\rm B})\Phi
-\alpha_{\rm B} \Psi+
\left( 2\alpha_{\rm B} \Delta_1-\alpha_{\rm B}^2+\Delta_2 \right)
\frac{H \delta \phi_{\rm N}}{\dot{\phi}} \right]
+(1-\beta_{n_c}-q_c ) \rho_c \dot{\delta}_{c{\rm N}}
-H \rho_c \beta_{n_c}\epsilon_{\beta_{n_c}} \delta_{c{\rm N}}=0\,.
\ea
Solving these equations and Eq.~(\ref{aniso}) 
for $\Psi$, $\Phi$, and $\delta \phi_{\rm N}$, 
we obtain
\ba
\hspace{-0.8cm}
\Psi &=& -\frac{a^2}{2q_t \Delta_2 k^2} \left[ 
\{\Delta_1\Delta_3+(1-\beta_{n_c})\Delta_2\} \rho_c \delta_{c{\rm N}}
+ (\Delta_1^2+\Delta_2)\rho_b \delta_{b{\rm N}}
+(1-\beta_{n_c}-q_c) 
\Delta_1 \rho_c \frac{\dot{\delta}_{c{\rm N}}}{H}\right],
\label{Psif} \\
\hspace{-0.8cm}
\Phi &=& \frac{a^2}{2q_t \Delta_2 k^2} \left[ 
\{\alpha_{\rm B}\Delta_3+(1-\beta_{n_c})\Delta_2\} \rho_c \delta_{c{\rm N}}
+(\alpha_{\rm B}\Delta_1+\Delta_2) \rho_b \delta_{b{\rm N}}
+(1-\beta_{n_c}-q_c) \alpha_{\rm B} \rho_c \frac{\dot{\delta}_{c{\rm N}}}{H}\right],
\label{Phif}\\
\delta \phi_{\rm N} &=&
-\frac{a^2 \dot{\phi}}{2H q_t \Delta_2 k^2} \left[
\Delta_3 \rho_c \delta_{c{\rm N}}
+\Delta_1 \rho_b \delta_{b{\rm N}}
+(1-\beta_{n_c}-q_c) \rho_c \frac{\dot{\delta}_{c{\rm N}}}{H}
\right]\,.
\label{delphif} 
\ea

Under the quasi-static approximation, the terms on the right-hand-side 
of Eq.~(\ref{quasi0}) can be neglected relative to those on the left-hand-side.
Substituting Eq.~(\ref{delphif}) and its time derivative as well as Eq.~(\ref{Psif}) 
into the left-hand-side of Eq.~(\ref{quasi0}), we obtain
\be
\ddot{\delta}_{c{\rm N}}+c_1 H \dot{\delta}_{c{\rm N}}
+c_2 H \dot{\delta}_{b{\rm N}}-\frac{3H^2}{2G} \left(G_{cc}\Omega_c\delta_{c{\rm N}}
+G_{cb}\Omega_b\delta_{b{\rm N}}\right)=0\,, 
\label{ddotceqf}
\ee
where $G=1/(8\pi M_{\rm pl}^2)$ is the Newton gravitational constant, and
\ba
c_1 &=& \left( 2+\epsilon_{q_c} \right) \frac{\hat{c}_s^2}{c_s^2}+
\left[ \frac{2\Delta_3-2q_c (\Delta_1+\epsilon_{q_c})}{1-\beta_{n_c}-q_c}
-1-\Delta_1-\alpha_{\rm B}-\epsilon_{\Delta_2}-2\epsilon_{\rm H} 
\right] \left( 1-\frac{\hat{c}_s^2}{c_s^2} \right)\,,\\
c_2 &=& \frac{3(1-\beta_{n_c}-q_c) \Omega_b \Delta_1}{2Q_t q_c \Delta_2}
\frac{\hat{c}_s^2}{c_s^2}\,,\\
G_{cc} &=& \frac{\Delta_1 \Delta_3 q_c+\Delta_2(1-\beta_{n_c})^2
+\Delta_3[q_c\epsilon_{q_c}+(1-\beta_{n_c}-q_c)
(1+\alpha_{\rm B}+\epsilon_{\rm H}
+\epsilon_{\Delta_2}-\epsilon_{\Delta_3})]}{q_c Q_t \Delta_2} 
\frac{\hat{c}_s^2}{c_s^2}G\,,\label{Gcc} \\
G_{cb} &=&\frac{\Delta_1^2 q_c+\Delta_2(1-\beta_{n_c})
+\Delta_1[q_c \epsilon_{q_c}+(1-\beta_{n_c}-q_c)
(1+\alpha_{\rm B}+\epsilon_{\rm H}-\epsilon_{\Delta_1}+\epsilon_{\Delta_2})]}
{q_c Q_t \Delta_2} 
\frac{\hat{c}_s^2}{c_s^2}G\,,\label{Gcb}
\ea
with
\be
Q_t=\frac{q_t}{M_{\rm pl}^2}\,,\qquad 
\epsilon_{\Delta_i}=\frac{\dot{\Delta}_i}
{H \Delta_i}\,,\qquad {\rm for}\quad i=1,2,3\,.
\ee
The CDM and baryon density parameters are defined 
in Eq.~(\ref{Omedef}), i.e., 
$\Omega_c=\rho_c/(3 M_{\rm pl}^2 H^2)$ and 
$\Omega_b=\rho_b/(3 M_{\rm pl}^2 H^2)$. 
The relation between $c_s^2$ and $\hat{c}_s^2$ is given by 
\be
c_s^2=\hat{c}_s^2 \left[ 1+\frac{3\Omega_c (1-\beta_{n_c}-q_c)^2}
{2q_c Q_t \Delta_2} \right]\,.
\ee
In Eq.~(\ref{ddotceqf}), there is no effective pressure of the form 
$c_{\rm CDM}^2(k^2/a^2) \delta_{c{\rm N}}$ as expected.
This property is attributed to the assumption (\ref{fnn}) 
of the coupling $f$ as well as the vanishing value of $c_c^2$. 
The clustering of CDM density perturbations occurs by the gravitational couplings
$G_{cc}$ and $G_{cb}$, both of which generally differ from $G$.

The baryon density contrast $\delta_{b{\rm N}}$ and velocity potential 
$v_{b{\rm N}}$ obey
\ba
& &
\dot{\delta}_{b{\rm N}}+3\dot{\Phi}+\frac{k^2}{a^2}v_{b{\rm N}}=0\,,\\
& &
\dot{v}_{b{\rm N}}-\Psi=0\,,
\ea
so that the second-order equation for $\delta_{b{\rm N}}$ is
\be
\ddot{\delta}_{b{\rm N}}+2H \dot{\delta}_{b{\rm N}}+\frac{k^2}{a^2} 
\Psi=-3\ddot{\Phi}-6H\dot{\Phi}\,.
\label{ddotbeq}
\ee
Neglecting the right-hand-side of Eq.~(\ref{ddotbeq}) and 
substituting Eq.~(\ref{Psif}) into Eq.~(\ref{ddotbeq}), it follows that 
\be
\ddot{\delta}_{b{\rm N}}+2H \dot{\delta}_{b{\rm N}}
-\frac{3\Delta_1 (1-\beta_{n_c}-q_c)\Omega_c}{2 Q_t \Delta_2} H \dot{\delta}_{c{\rm N}}
-\frac{3H^2}{2G} \left(G_{bc}\Omega_c\delta_{c{\rm N}}
+G_{bb}\Omega_b\delta_{b{\rm N}}\right)=0\,, 
\label{ddotbeqf}
\ee
where 
\ba
G_{bc} &=& \frac{\Delta_1\Delta_3+(1-\beta_{n_c})\Delta_2}{Q_t \Delta_2}G\,,
\label{Gbc} \\
G_{bb} &=& \frac{\Delta_1^2+\Delta_2}{Q_t \Delta_2}G\,.
\label{Gbb}
\ea

The baryon density perturbation is directly affected by 
the evolution of gravitational potential $\Psi$. 
The difference from uncoupled Horndeski theories is that the time derivative 
$\dot{\delta}_{c{\rm N}}$ appears in the expression of $\Psi$ 
given by Eq.~(\ref{Psif}). 
By defining the dimensionless quantities, 
\be
f_c=\frac{\dot{\delta}_{c{\rm N}}}{H \delta_{c{\rm N}}}\,,\qquad 
\mu_{bc}=\frac{G_{bc}}{G}\,,\qquad
\mu_{bb}=\frac{G_{bb}}{G}\,,
\ee
one can express Eq.~(\ref{Psif}) in the form of Poisson equation, 
\be
\frac{k^2}{a^2} \Psi=-4\pi G \left[ \left\{ \mu_{bc}
+\frac{f_c(1-\beta_{n_c}-q_c) \Delta_1}{Q_t \Delta_2}
\right\} \rho_c \delta_{c{\rm N}}+\mu_{bb} \rho_b \delta_{b{\rm N}}
\right]\,.
\label{Poisson}
\ee
The gravitational potential associated with the observations of weak lensing 
is defined by \cite{Sapone}
\be
\psi_{\rm WL}=\frac{1}{2} \left(\Psi-\Phi \right)\,.
\ee
On using Eqs.~(\ref{Psif}) and (\ref{Phif}) together with the relation 
$\alpha_{\rm B}=\Delta_1+\alpha_{\rm M}$, it follows that 
\be
\frac{k^2}{a^2} \psi_{\rm WL}=-4\pi G \left[ \left\{ \mu_{bc}
+\frac{\alpha_{\rm M} \Delta_3 +f_c(1-\beta_{n_c}-q_c)
(2\Delta_1+\alpha_{\rm M})}{2Q_t \Delta_2}
\right\} \rho_c \delta_{c{\rm N}}
+\left( \mu_{bb}+\frac{\alpha_{\rm M} \Delta_1}{2Q_t \Delta_2} 
\right) \rho_b \delta_{b{\rm N}}
\right]\,.
\label{psiWL}
\ee
If $\alpha_{\rm M}=0$, then the right-hand-sides of Eqs.~(\ref{Poisson}) 
and (\ref{psiWL}) coincide with each other, 
so that $\psi_{\rm WL}=\Psi=-\Phi$.
This is the consequence of 
the absence of anisotropic stress in Eq.~(\ref{aniso}). 
If the anisotropic stress is present, there are contributions 
to $\psi_{\rm WL}$ arising from a nonvanishing value of $\alpha_{\rm M}$.
The CDM growth rate $f_c$ also appears on the right-hand-side 
of Eq.~(\ref{psiWL}). The dynamics of $\delta_{c{\rm N}}$ and 
$\delta_{b{\rm N}}$ for perturbations deep inside the 
sound horizon is known by solving Eqs.~(\ref{ddotceqf}) 
and (\ref{ddotbeqf}) with the gravitational couplings 
(\ref{Gcc})-(\ref{Gcb}) and (\ref{Gbc})-(\ref{Gbb}). 
The modified evolution of $\delta_{c{\rm N}}$ and $\delta_{b{\rm N}}$ 
in comparison to the theories with $f=0$ affects the dynamics of 
gravitational potentials $\Psi$ and $\psi_{\rm WL}$ through 
Eqs.~(\ref{Poisson}) and (\ref{psiWL}).

The effect of DE and DM interactions 
on $G_{cc}$, $G_{cb}$, and $G_{bc}$ 
appears through the two quantities $q_c$ and $\beta_{n_c}$. 
The $Z$ and $n_c$ dependence in $q_c$ leads to 
the deviation of $q_c$ from 1.
If there is no $n_c$ dependence in $f$, the quantity $\beta_{n_c}$ 
vanishes. This means that the deviation of $\beta_{n_c}$ from 0 occurs through 
the energy transfer associated with the change of $n_c$. 
Now, we are considering the interacting theories satisfying the condition 
(\ref{fnn}), under which the coupling $f$ is constrained to be 
of the form (\ref{fcon2}) with the linear dependence 
$\rho_c \propto n_c$. 
In this case, we have
\ba
q_c &=& 1+f_1-\dot{\phi} f_{1,Z}
+\frac{\dot{\phi} f_{2,Z}}{\rho_c} \,,\label{qcf} \\
\beta_{n_c} &=& -f_1+\dot{\phi}^2 f_{1,X}
+\dot{\phi}f_{1,Z}\,.
\label{betacf}
\ea
This means that $\beta_{n_c}$ depends on $f_1$ alone, 
while $q_c$ contains the dependence of both $f_1$ and $f_2$.
For the theories with $f_1=0$, we have 
$q_c=1+\dot{\phi}f_{2,Z}/\rho_c$ and $\beta_{n_c}=0$.
In this case, the interaction between CDM and $\phi$ 
occurs through the momentum transfer characterized 
by the $Z$ dependence in $f_2$. 
In the limit that $q_c \to 1$, $\epsilon_{q_c} \to 0$, $\beta_{n_c} \to 0$, 
and $\epsilon_{\beta_{n_c}} \to 0$, one can conform that 
$G_{cc}$, $G_{cb}$, and $G_{bc}$ reduce to $G_{bb}$ given 
by Eq.~(\ref{Gbb}).
This value of $G_{bb}$ is identical to the gravitational coupling
of baryons and CDM derived for uncoupled Horndeski 
theories \cite{DeFelice:2011hq,Kase:2018aps}, 
which is larger than 
$G/Q_t$ under the absence of ghosts and Laplacian instabilities. 
The existence of coupling $f$ generally leads to the values of 
$G_{cc}$, $G_{cb}$, and $G_{bc}$ different from $G_{bb}$.

\section{Gravitational couplings in concrete theories}
\label{conthesec}

In concrete interacting theories of DE and DM, we compute the gravitational couplings 
$G_{cc}$, $G_{cb}$, $G_{bc}$, and $G_{bb}$ derived in Sec.~\ref{Geffsec}. 
For this purpose, we will focus on the coupling function of the form, 
\be
f=-f_1(\phi,X,Z) \rho_c+f_2(\phi,X,Z)\,,
\label{fcon3}
\ee
which satisfies the condition $f_{,n_c n_c}=0$.
We classify the theories into two classes: 
(i) $f_1=0$, and (ii) $f_1 \neq 0$. 
In each class, we estimate the values of $G_{cc}$, $G_{cb}$, 
$G_{bc}$, and $G_{bb}$ for theories which belong to 
the coupling (\ref{fcon3}).

\subsection{$f_1=0$ and $f_2 \neq 0$}

We begin with interacting theories in which 
the coupling $f_1$ is absent, i.e., 
\be
f=f_2(\phi, X, Z)\,.
\ee
In this case, the quantities (\ref{qcf}) and (\ref{betacf}) reduce, 
respectively, to  
\be
q_c=1+\frac{\dot{\phi} f_{2,Z}}{\rho_c}\,,\qquad
\beta_{n_c} = 0\,.
\ee
With this latter relation, we have
\be
\Delta_3=\Delta_1\,,\qquad 
\epsilon_{\Delta_3}=\epsilon_{\Delta_1}\,.
\ee
Then, the gravitational couplings (\ref{Gcc})-(\ref{Gcb}) 
and (\ref{Gbc})-(\ref{Gbb}) yield
\ba
G_{cc} &=& G_{cb}=
\frac{\Delta_1^2 q_c+\Delta_2+\Delta_1 q_c \epsilon_{q_c}-\Delta_1(q_c-1)
(1+\alpha_{\rm B}+\epsilon_{\rm H}-\epsilon_{\Delta_1}+\epsilon_{\Delta_2})}
{q_c Q_t \Delta_2} \frac{\hat{c}_s^2}{c_s^2}G\,,\label{Gcc0} \\
G_{bb} &=& G_{bc}=\frac{\Delta_1^2+\Delta_2}{Q_t \Delta_2}G\,,
\label{Gbc0}
\ea
where 
\be
\frac{c_s^2}{\hat{c}_s^2}=1+\frac{3\Omega_c (1-q_c)^2}
{2q_c Q_t \Delta_2}\,.
\label{csr}
\ee
Hence $G_{cc}$ and $G_{bb}$ are equivalent to $G_{cb}$ 
and $G_{bc}$, respectively.
In the limit that $q_c \to 1$ and $\epsilon_{q_c} \to 0$, 
$G_{cc}$ reduces to $G_{bb}$.

The quantity $\Delta_1=\alpha_{\rm B}-\alpha_{\rm M}$ is different 
depending on the choice of Horndeski Lagrangian (\ref{LH}). 
In the following, we will consider three different cases: 
(a) k-essence, (b) extended Galileons, and (c) nonminimal couplings.

\subsubsection{k-essence}
\label{kessec}

Let us first consider minimally coupled k-essence 
theories \cite{kes1,kes2,kes3}
given by the Lagrangian  
\be
{\cal L}_{\rm H}=\frac{M_{\rm pl}^2}{2}R+G_2(\phi,X)\,.
\ee
Since $\alpha_{\rm B}=0$ and $\alpha_{\rm M}=0$ 
in this case, it follows that 
\be
\Delta_1=0\,,
\ee
with $Q_t=1$.
Then, Eqs.~(\ref{Gcc0}) and (\ref{Gbc0}) reduce, respectively, to 
\ba
G_{cc} &=& G_{cb}=\frac{\hat{c}_s^2}{q_c c_s^2}G\,,
\label{Gcce} \\
G_{bb} &=& G_{bc}=G\,.
\ea
The baryon gravitational couplings $G_{bb}$ and $G_{bc}$ 
are equivalent to the Newton constant $G$, but 
$G_{cc}$ and $G_{cb}$ are different from $G$.
Since 
\be
\Delta_2=\frac{\dot{\phi}^2 (G_{2,X}+f_{2,X})}{2M_{\rm pl}^2 H^2}\,,
\ee
the ratio (\ref{csr}) is expressed as 
\be
\frac{c_s^2}{\hat{c}_s^2}=1+\frac{f_{2,Z}^2}
{(G_{2,X}+f_{2,X})(\dot{\phi}f_{2,Z}+\rho_c)}\,,
\ee
where we used $\rho_c=3M_{\rm pl}^2 H^2 \Omega_c$ 
instead of $\Omega_c$.
Then, from Eq.~(\ref{Gcce}), we obtain 
\be
G_{cc}=G_{cb}=\frac{G}{1+r_{f_2}}\,,
\label{Gccrf}
\ee
where 
\be
r_{f_2}=\frac{(G_{2,X}+f_{2,X})\dot{\phi} f_{2,Z}+f_{2,Z}^2}
{(G_{2,X}+f_{2,X})\rho_c}\,.
\label{rf2}
\ee
In quintessence given by the Lagrangian $G_2(\phi, X)=X-V(\phi)$, 
we have $G_{2,X}=1$ in Eq.~(\ref{rf2}), which coincides with 
the result derived in Ref.~\cite{Kase:2019mox}.
Now, we showed that the generalized expression (\ref{Gccrf}) 
with (\ref{rf2}) holds for k-essence.

Provided that $\rho_c$ dominates over
the density associated with the coupling $f_2$ in the early 
matter era, we have $r_{f_2} \ll 1$ and hence $G_{cc} \simeq G$. 
The deviation of $G_{cc}$ from $G$ starts to occur after the 
dominance of DE, around which the term 
$(G_{2,X}+f_{2,X})\dot{\phi} f_{2,Z}+f_{2,Z}^2$ becomes the 
same order as $(G_{2,X}+f_{2,X})\rho_c$.
In this epoch, the CDM gravitational interaction weaker than 
the Newton constant $G$ can be realized for $r_{f_2}>0$. 
In the DE sector, the ghost and Laplacian instabilities 
are absent for
\ba
q_s &=& 2M_{\rm pl}^2 \left( G_{2,X}+\dot{\phi}^2 G_{2,XX}
+f_{2,X}+\dot{\phi}^2 f_{2,XX}
+f_{2,ZZ}+2\dot{\phi} f_{2,XZ} \right)>0\,,\label{qsf2} \\
q_s\hat{c}_s^2 
&=& 2M_{\rm pl}^2 \left( G_{2,X}+f_{2,X} 
\right)>0\,.\label{csf2} 
\ea
Under the requirement (\ref{csf2}), the condition $G_{cc}<G$ 
translates to
\be
(G_{2,X}+f_{2,X})\dot{\phi} f_{2,Z}+f_{2,Z}^2>0\,,
\label{weakcon}
\ee
which is satisfied for $\dot{\phi}f_{2,Z}>0$. 
If we consider the interaction 
$f_2=\beta Z^2$ \cite{Pourtsidou:2013nha,Pourtsidou:2016ico,Kase:2019mox}, 
for example, the positive coupling constant $\beta$ always leads to the weak 
CDM gravitational interaction.

The evolution of $G_{cc}$ in the asymptotic future depends on 
the scalar time-derivative $\dot{\phi}$. 
For quintessence with an exponential potential, i.e., 
$G_2=X-V_0 e^{-\lambda \phi/M_{\rm pl}}$ with $\lambda^2<2$, 
there exists a future accelerating fixed point along which 
$\dot{\phi}$ is proportional to $H$, with the DE 
equation of state $w_{\rm DE}
=-1+\lambda^2/3$ and density parameter $\Omega_{\rm DE}=1$ \cite{CLW,CST}. 
In this case, the $\dot{\phi}$-dependent terms in $r_{f_2}$ slowly 
decrease in comparison to $\rho_c~(\propto a^{-3})$ 
and hence $r_{f_2}$ grows continuously.
Then, $G_{cc}$ and $G_{cb}$ approach 0 toward the future 
accelerating fixed point \cite{Kase:2019mox}. 
This is also the case for k-essence allowing for the existence of 
a future de Sitter solution characterized by 
$\dot{\phi}={\rm constant}$ and $H={\rm constant}$. 
We also note that the quantity $q_c$ asymptotically behaves as 
$q_c \simeq \dot{\phi}f_{2,Z}/\rho_c$, so the absence of ghosts in the 
CDM sector requires that $\dot{\phi}f_{2,Z}>0$. 
In this case the condition (\ref{weakcon}) is satisfied, so the weak 
gravitational interaction for CDM is naturally realized 
after the dominance of DE.

In summary, k-essence with the $Z$-dependent contributions to $f_2$ 
leads to $G_{cc}$ smaller than $G$ at low redshifts, 
with $G_{cc}$ approaching 0 in the future.

\subsubsection{Extended Galileons}

The minimally coupled extended Galileon is given by 
the Lagrangian  
\be
{\cal L}_{\rm H}=\frac{M_{\rm pl}^2}{2}R
+G_2(X)+G_3(X) \square \phi\,,
\label{LHga}
\ee
where $G_2$ and $G_3$ depend on $X$ alone. 
The cubic Galileon \cite{Nicolis,Deffayet1} is characterized by 
the functions  $G_2(X)=c_2 X$ and $G_3(X)=c_3 X$, 
where $c_2$ and $c_3$ are constants. 
The extended Galileon \cite{Deffayet2,DeFelice:2011bh} corresponds to the theories 
with arbitrary functions of $G_2(X)$ and $G_3(X)$.

For the Lagrangian (\ref{LHga}), we have 
\be
\alpha_{\rm M}=0\,,\qquad \Delta_1=
\alpha_{\rm B}=\frac{\dot{\phi}^3 G_{3,X}}{2H M_{\rm pl}^2} \,,
\ee
and $Q_t=1$. Then, the gravitational couplings (\ref{Gcc0}) 
and (\ref{Gbc0}) reduce to 
\ba
G_{cc} &=& G_{cb}=
\frac{\alpha_{\rm B}^2+\Delta_2+\alpha_{\rm B}q_c \epsilon_{q_c}
-\alpha_{\rm B} (q_c-1)
(1+\epsilon_{\rm H}-\epsilon_{\Delta_1}+\epsilon_{\Delta_2})}
{q_c \Delta_2} \frac{\hat{c}_s^2}{c_s^2}G\,,\label{Gccga} \\
G_{bb} &=& G_{bc}=\frac{\alpha_{\rm B}^2+\Delta_2}{\Delta_2}G\,,
\label{Gbcga}
\ea
where 
\ba
\Delta_2 &=& 
\frac{\dot{\phi}^2[2M_{\rm pl}^2(f_{2,X}+G_{2,X}-2 \ddot{\phi} G_{3,X}
-4H \dot{\phi} G_{3,X}-\ddot{\phi} \dot{\phi}^2 G_{3,XX})
-\dot{\phi}^4 G_{3,X}^2]}{4H^2 M_{\rm pl}^4}\,,\label{Deltaga}\\
\frac{c_s^2}{\hat{c}_s^2} &=& 1+\frac{\dot{\phi}^2 f_{2,Z}^2}
{2H^2 M_{\rm pl}^2 (\dot{\phi} f_{2,Z}+\rho_c)\Delta_2}\,.
\label{csga}
\ea
Since $\Delta_2>0$ for the absence of ghosts and Laplacian instabilities, 
$G_{bb}$ and $G_{bc}$ are larger than $G$. 
The braiding term $\alpha_{\rm B}^2$, which arises from the cubic 
coupling $G_3(X)$, leads to the enhancement of baryon gravitational couplings. 
In contrast, the CDM perturbation is affected not only by the term 
$\alpha_{\rm B}^2$ but also by the deviation 
of $q_c$ from 1 (which is induced by the $Z$-dependence in $f_2$). 
The latter term allows a possibility for realizing 
$G_{cc}$ smaller than $G$.

Subtracting Eq.~(\ref{back1}) from Eq.~(\ref{back2}), we obtain
\be
-2M_{\rm pl}^2 \dot{H}=\dot{\phi} \left( \dot{\phi} f_{2,X}+ f_{2,Z}
+\dot{\phi} G_{2,X}-3H \dot{\phi}^2 G_{3,X}+\dot{\phi} \ddot{\phi} 
G_{3,X} \right)+\sum_{I=c,b} \left( \rho_I+P_I \right)\,.
\ee
Provided that $f_2$ does not contain the $\phi$ dependence, i.e., 
\be
f=f_2(X, Z)\,, 
\ee
there exists a de Sitter solution characterized by 
\be
H={\rm constant}\,,\qquad 
\dot{\phi}={\rm constant}, \qquad 
\rho_I=0=P_I\,.
\label{deSitter}
\ee
Along this solution, the nonvanishing time derivative 
$\dot{\phi}$ obeys
\be
f_{2,Z}+\dot{\phi} \left( f_{2,X}
+G_{2,X}-3H \dot{\phi} G_{3,X} \right)=0\,.
\ee
Then, the quantities (\ref{Deltaga}) and (\ref{csga}) reduce, respectively, to 
\ba
\Delta_2 &=& -\alpha_{\rm B} \left( \alpha_{\rm B}+1 \right)-g_2\,,\\
\frac{c_s^2}{\hat{c}_s^2} &=& \frac{\alpha_{\rm B} \left( \alpha_{\rm B}+1 \right)}
{\alpha_{\rm B} \left( \alpha_{\rm B}+1 \right)+g_2}\,,
\ea
where 
\be
g_2=\frac{\dot{\phi}f_{2,Z}}{2H^2 M_{\rm pl}^2}\,.
\ee
Requiring that the quantity $q_c$ around 
the de Sitter solution ($q_c \simeq \dot{\phi}f_{2,Z}/\rho_c$) is positive, 
it follows that $g_2>0$. Since $\Delta_2>0$ to avoid the ghost and 
Laplacian instabilities, it is at least necessary to 
satisfy the inequality $-\alpha_{\rm B} \left( \alpha_{\rm B}+1 \right)>0$, i.e., 
\be
-1<\alpha_{\rm B}<0\,.
\label{aBcon}
\ee
On the de Sitter fixed point characterized by Eq.~(\ref{deSitter}), 
the quantities appearing in Eq.~(\ref{Gcc0}) satisfy 
\be
\epsilon_{q_c}=3\,,\qquad \epsilon_{\rm H}=0\,, \qquad 
\epsilon_{\Delta_1}=0\,,\qquad 
\epsilon_{\Delta_2}=0\,.
\ee
Taking the limit $q_c \to \infty$ in Eq.~(\ref{Gcc0}), the 
CDM gravitational coupling on the de Sitter solution yields
\be
(G_{cc})_{\rm dS}=\frac{2\alpha_{\rm B}}{\Delta_2}\frac{\hat{c}_s^2}{c_s^2}G
=-\frac{2}{\alpha_{\rm B}+1}G\,.
\label{GccdS}
\ee
Since $(G_{cc})_{\rm dS}$ is negative under the condition (\ref{aBcon}),
the CDM gravitational interaction is repulsive. 
This peculiar behavior results from the momentum exchange 
between CDM and the self-accelerating scalar field. 
As we observe in Eq.~(\ref{Gccga}), the terms $\alpha_{\rm B}^2+\Delta_2$, 
which also appear in the numerator of Eq.~(\ref{Gbcga}), are completely 
dominated by the $q_c$-dependent contributions to $G_{cc}$ 
on the de Sitter solution. This means that the $Z$-dependence in $f_2$ 
gives the value of $(G_{cc})_{\rm dS}$ very different from $(G_{bb})_{\rm dS}$. 
We also note that the result (\ref{GccdS}) agrees with the weak 
vector-field coupling limit of the CDM gravitational coupling derived 
for generalized Proca theories \cite{DeFelice:2020icf} (with the change of notation 
$\alpha_{\rm B} \to -\alpha_{\rm B}$).

Provided that $q_c \simeq 1$ in the early matter 
era, $G_{cc}$ is close to the value $G_{bb}~(>G)$.
The evolution of $G_{cc}$ just after the dominance of DE 
depends on the forms of $f_2$ and $G_3$. 
Since $(G_{cc})_{\rm dS}<0$, the CDM perturbation should eventually 
cross the point $(G_{cc})_{\rm dS}=0$ on the way of approaching 
the future de Sitter fixed point.
The moment at which this transition occurs depends on the 
chosen model parameters.

\subsubsection{Nonminimal couplings}

We proceed to nonminimally coupled k-essence theories 
characterized by the Lagrangian 
\be
{\cal L}_{\rm H}=G_4(\phi)R+G_2(\phi,X)\,.
\ee
In this case, we have 
\be
\alpha_{\rm M}=2\alpha_{\rm B}=\frac{\dot{\phi} G_{4,\phi}}{H G_4}\,,
\qquad 
\Delta_1=-\alpha_{\rm B}=
-\frac{\dot{\phi} G_{4,\phi}}{2H G_4}\,,\qquad 
Q_t=\frac{2G_4}{M_{\rm pl}^2}\,.
\ee
Then, the gravitational couplings are given by 
\ba
G_{cc} &=& G_{cb}=
\frac{(2q_c-1)\alpha_{\rm B}^2+\Delta_2-\alpha_{\rm B}q_c \epsilon_{q_c}
+\alpha_{\rm B} (q_c-1)
(1+\epsilon_{\rm H}-\epsilon_{\Delta_1}+\epsilon_{\Delta_2})}
{q_c Q_t \Delta_2} \frac{\hat{c}_s^2}{c_s^2}G\,,\label{Gccnon} \\
G_{bb} &=& G_{bc}=\frac{\alpha_{\rm B}^2+\Delta_2}{Q_t\Delta_2}G\,,
\label{Gbcnon}
\ea
where
\ba
\Delta_2 &=& \frac{\dot{\phi}^2 [G_4(G_{2,X}+f_{2,X})
+3G_{4,\phi}^2]}{4H^2 G_4^2}\,,\\
\frac{c_s^2}{\hat{c}_s^2} &=& 
1+\frac{G_4 f_{2,Z}^2}
{[G_4(G_{2,X}+f_{2,X})
+3G_{4,\phi}^2](\dot{\phi}f_{2,Z}+\rho_c)}\,.
\ea
Provided that $q_c$ and $c_s^2/\hat{c}_s^2$ are close to 1 during the 
early matter era due to the smallness of $f_{2,Z}$, $G_{cc}$ reduces to 
the value $G_{bb}$ in Eq.~(\ref{Gbcnon}).
As $q_c$ and $c_s^2/\hat{c}_s^2$ start to deviate from 1 
at low redshifts, $G_{cc}$ exhibits the different evolution from $G_{bb}$. 
If the term $\dot{\phi}f_{2,Z}$ decreases slowly in comparison to $\rho_c$ 
in the late Universe, then $q_c$ continuously grows toward infinity. 
Taking the limit $q_c \to \infty$ in Eq.~(\ref{Gccnon}), it follows that 
\be
(G_{cc})_{\rm late}=\frac{\alpha_{\rm B} 
(1-\epsilon_{q_c}+\epsilon_{\rm H}+2\alpha_{\rm B}
-\epsilon_{\Delta_1}+\epsilon_{\Delta_2})}{Q_t \Delta_2}
\frac{\hat{c}_s^2}{c_s^2}G\,.
\label{Gccnon2}
\ee
If the scalar field $\phi$ evolves slowly on a quasi de-Sitter background, 
the terms $\epsilon_{\rm H}$, $\alpha_{\rm B}$, $\epsilon_{\Delta_1}$, 
and $\epsilon_{\Delta_2}$ should be much smaller than 1, with $\epsilon_{q_c} \simeq 3$. 
In this case, Eq.~(\ref{Gccnon2}) approximately reduces to 
\be
(G_{cc})_{\rm late} \simeq -\frac{2\alpha_{\rm B}}
{Q_t\Delta_2}\frac{\hat{c}_s^2}{c_s^2}G\,.
\ee
The dominant contributions to $(G_{cc})_{\rm late}$ arise from the terms 
$-\alpha_{\rm B}q_c \epsilon_{q_c}$ and $\alpha_{\rm B} (q_c-1)$ in the 
numerator of Eq.~(\ref{Gccnon}). 
This means that the momentum transfer between CDM and the scalar field 
completely dominates over the terms associated with nonminimal couplings 
on the quasi de-Sitter background.

\subsection{$f_1 \neq 0$ and $f_2 \neq 0$}

Finally, we study the interacting theories in which the coupling $f_1$ 
is present besides $f_2$. 
We focus on the simple case in which $f_1$ depends on $\phi$ 
alone, i.e., 
\be
f=-f_1(\phi) \rho_c+f_2(\phi, X, Z)\,.
\ee
For the DE sector, we consider the minimally coupled k-essence 
given by the Lagrangian,
\be
{\cal L}_{\rm H}=\frac{M_{\rm pl}^2}{2}R+G_2(\phi,X)\,,
\ee
under which $\alpha_{\rm B}=0$, $\alpha_{\rm M}=0$, and $Q_t=1$.
In such theories, we have
\be
\Delta_1=0\,,\qquad 
\Delta_2=\frac{\dot{\phi}^2 (G_{2,X}+f_{2,X})}{2H^2 M_{\rm pl}^2}\,,\qquad 
\Delta_3=\frac{\dot{\phi} f_{1,\phi}}{H}\,,
\ee
and 
\be
q_c=1+f_1
+\frac{\dot{\phi} f_{2,Z}}{\rho_c}\,,\qquad 
\frac{c_s^2}{\hat{c}_s^2}=1+\frac{f_{2,Z}^2}
{(G_{2,X}+f_{2,X})[(1+f_1)\rho_c+\dot{\phi}f_{2,Z}]}\,.
\ee
Then, the gravitational couplings (\ref{Gcc})-(\ref{Gcb}) and (\ref{Gbc})-(\ref{Gbb}) 
are expressed as 
\ba
G_{cc} &=& \frac{1+f_1+r_1}{1+r_2}G\,,\label{Gccf1}\\
G_{cb} &=& \frac{1}{1+r_2}G\,,\label{Gcbf1}\\
G_{bc} &=& (1+f_1)G\,,\label{Gbcf1}\\
G_{bb} &=& G\,,
\ea
where 
\ba
r_1 &=& -\frac{2H M_{\rm pl}^2 f_{1,\phi}}{(G_{2,X}+f_{2,X})(1+f_1)\rho_c} 
\left[ f_{2, Z} \left( 1-\epsilon_{q_c}+\epsilon_{\rm H}+\epsilon_{\Delta_2}
-\epsilon_{\Delta_3} \right)-(1+f_1) \frac{\rho_c \epsilon_{q_c}}{\dot{\phi}} 
\right]\,,\label{r1non}\\
r_{2} &=& \frac{(G_{2,X}+f_{2,X})\dot{\phi} f_{2,Z}+f_{2,Z}^2}
{(G_{2,X}+f_{2,X})(1+f_1)\rho_c}\,.\label{r2non}
\ea
Taking the limit $f_1 \to 0$, these gravitational couplings coincide with those 
derived in Sec.~\ref{kessec}.
The nonvanishing function $f_1$ leads to the difference between $G_{cc}$ and $G_{cb}$, 
and also between $G_{bc}$ and $G_{bb}$. 
The $\phi$-dependence in $f_1$ gives rise to a new contribution $r_1$ 
to the numerator of $G_{cc}$ in Eq.~(\ref{Gccf1}). This contribution
arises through the energy exchange between the scalar field and CDM. 
The momentum transfer between $\phi$ and CDM, which appears 
as the $Z$ dependence in $f_2$, occurs 
through the term $r_2$ in the denominators of $G_{cc}$ and $G_{cb}$.
{}From Eq.~(\ref{Gcbf1}), we find that $G_{cb}$ is affected only 
by the momentum exchange.  
The coupling $f_1$ modifies the amplitude of $G_{bc}$, but 
$G_{bb}$ is equivalent to $G$.

As long as the conditions $|r_1| \ll 1$ and $|r_2| \ll 1$ 
are satisfied in the early matter era, $G_{cc}$ and $G_{cb}$ 
are close to $G_{bc}=(1+f_1)G$ and $G_{bb}=G$, respectively.
After the DE density dominates over the CDM density, 
$G_{cc}$ and $G_{cb}$ start to deviate from their initial values. 
Let us consider the case in which the scalar field 
evolves slowly after the dominance of DE. 
At a sufficiently late epoch in which $\rho_c$ becomes negligibly 
small relative to the DE density, one can take the limit 
$\rho_c \to 0$ in Eqs.~(\ref{Gccf1}) and (\ref{Gcbf1}), 
with Eqs.~(\ref{r1non}) and (\ref{r2non}).
In this regime, the CDM gravitational couplings reduce to
\ba
(G_{cc})_{\rm late} &\simeq& -\frac{2f_{1,\phi}H M_{\rm pl}^2
(1-\epsilon_{q_c}+\epsilon_{\rm H}+\epsilon_{\Delta_2}-\epsilon_{\Delta_3})}
{(G_{2,X}+f_{2,X})\dot{\phi}+f_{2,Z}}G\,,\label{Gccqds}\\
(G_{cb})_{\rm late} &\simeq& 0\,.\label{Gcbqds}
\ea
The $\phi$-dependence in $f_1$ renders $(G_{cc})_{\rm late}$ 
different from 0, while $G_{cb}$ asymptotically approaches 0.

For concreteness, let us consider the coupling function, 
\be
f=-\left( e^{Q \phi/M_{\rm pl}}-1 \right) \rho_c
+\beta \left( 2X \right)^{1-m/2} Z^m\,,
\label{model2}
\ee
and quintessence with an exponential potential,
\be
G_2=X-V_0 e^{-\lambda \phi/M_{\rm pl}}\,,
\ee
where $Q, \beta, m, V_0, \lambda$ are constants. 
In this model, there exists the scalar-field dominated fixed 
point satisfying \cite{Amendola:2020ldb}
\be
\frac{\dot{\phi}}{H M_{\rm pl}}=\frac{\lambda}{1+2\beta}\,,\qquad
\epsilon_{\rm H}=-\frac{\lambda^2}{2(1+2\beta)}\,,\qquad 
\Omega_c=0\,,
\ee
at which we have
\be
\epsilon_{q_c}=3-\frac{\lambda^2}{1+2\beta}\,,\qquad
\epsilon_{\Delta_2}=0\,,\qquad 
\epsilon_{\Delta_3}=\frac{Q \lambda}{1+2\beta}\,.
\ee
Then, Eq.~(\ref{Gccqds}) reduces to
\be
(G_{cc})_{\rm late}=\frac{4(1+2\beta)+\lambda (2Q-\lambda)}{1+2\beta}
\frac{Q}{\lambda} e^{Q \phi/M_{\rm pl}} G\,,
\label{Gccasym}
\ee
which coincides with that derived in 
Ref.~\cite{Amendola:2020ldb}\footnote{In Ref.~\cite{Amendola:2020ldb}, 
the factor $e^{Q \phi/M_{\rm pl}}$ is absorbed into the definition of $\Omega_c$ 
in Eq.~(\ref{ddotceqf}), such that $\Omega_c \to (1+f_1)\Omega_c$.}. 
For $Q$ close to 0, $(G_{cc})_{\rm late}$ can be much smaller than $G$. 
In Ref.~\cite{Amendola:2020ldb}, it was shown that $G_{cc}$ can enter 
the region $G_{cc}<G$ by today and it finally approaches the asymptotic 
value (\ref{Gccasym}). Thus, even in the presence of the energy transfer arising from 
the coupling $f_1(\phi) \rho_c$, there are models in which the realization of 
weak cosmic growth at low redshifts is possible.

The CDM gravitational couplings (\ref{Gccf1})-(\ref{Gcbf1}) and their asymptotic values 
(\ref{Gccqds})-(\ref{Gcbqds}) can be applicable to arbitrary functions
$f_1(\phi)$, $f_2(\phi,X,Z)$ and the k-essence Lagrangian $G_2(\phi,X)$. 
Further extensions to theories with the coupling $f_1(\phi, X, Z)\rho_c$ and 
the Horndeski Lagrangian (\ref{LH}) are straightforward by using our
 most general formulas (\ref{Gcc})-(\ref{Gcb}) 
and (\ref{Gbc})-(\ref{Gbb}) of CDM and baryon gravitational couplings.

\begin{figure}[h]
\begin{center}
\includegraphics[height=3.2in,width=3.3in]{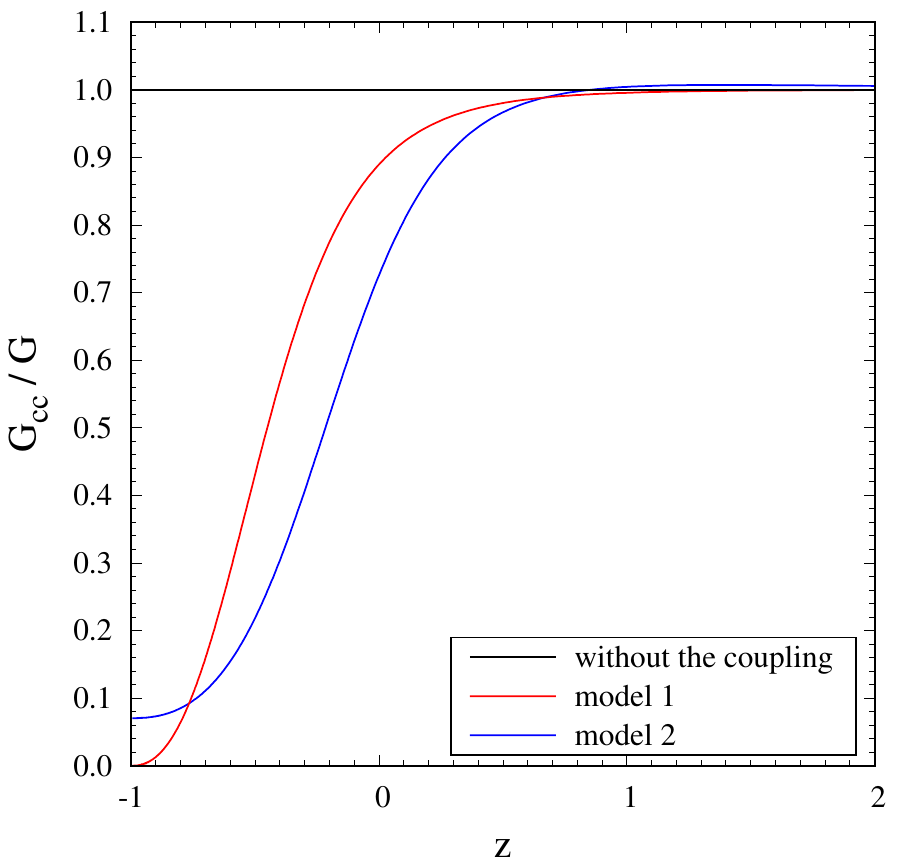}
\end{center}\vspace{-.5cm}
\caption{\label{fig1} 
Evolution of $G_{cc}$ versus the redshift $z$ in three different models. 
The red line corresponds to the model 1 with $\lambda=1$ and $\beta=1/4$, 
while the blue line to the model 2 with $\lambda=1$, $\beta=0.5$, and $Q=0.02$.
We also plot the case without the coupling $f$, i.e., $G_{cc}=G$, 
as a black line.
}
\end{figure}

Before closing this section, we explicitly show how the weak gravitational 
CDM coupling can be realized in concrete interacting theories.
In Fig.~\ref{fig1}, we plot the evolution of $G_{cc}$ versus the redshift $z$ 
for two concrete models 1 and 2 presented in Refs.~\cite{Kase:2019mox} 
and \cite{Amendola:2020ldb}, respectively.
The model 1 corresponds to the coupling $f_1=0$ and $f_2=\beta Z^2$, 
so it belongs to the class (A). The model 2, which belongs to the class (B), 
is characterized by the coupling (\ref{model2}) with $m=3$. 
In both models the Horndeski functions are 
given by $G_2=X-V_0e^{-\lambda\phi/M_{\rm pl}}$,
$G_3=0$, and $G_4=M_{\rm pl}^2/2$, 
with the model parameters $\lambda=1$, $\beta=1/4$ (model 1) and 
$\lambda=1$, $\beta=0.5$, and $Q=0.02$ (model 2). 
In Fig.~\ref{fig1}, we observe that $G_{cc}$ is smaller than $G$ 
at low redshifts, so the weak CDM gravitational interaction 
is indeed realized in these models.
In models 1 and 2, the asymptotic values of $G_{cc}$ in future 
are 0 and (\ref{Gccasym}), respectively, which are confirmed 
numerically.

\section{Conclusions}
\label{conclusion}

We studied very general interacting theories of DE and DM given by the action 
(\ref{action}), by paying particular attention to the gravitational couplings of 
CDM and baryon density perturbations. 
The DE sector is described by a scalar field $\phi$ with 
the Horndeski Lagrangian (\ref{LH}), whereas 
the CDM and baryons are dealt as perfect fluids characterized by the second integral 
in Eq.~(\ref{action}). The Lagrangian $f(n_c, \phi, X, Z)$ accommodates 
the interaction between DE and CDM. In particular, the $Z$ dependence 
in $f$ mediates the momentum transfer besides the energy exchange 
associated with the CDM number density $n_c$ coupled to the scalar field.

In Sec.~\ref{eomsec}, we derived the gravitational and CDM equations 
of motion in the covariant forms (\ref{ein}) and (\ref{ucov}). 
On the flat FLRW background, they reduce to Eqs.~(\ref{back1}) and (\ref{back2}), 
with the continuity Eqs.~(\ref{coneq}) and (\ref{coneq2}). 
If we consider the interacting Lagrangian (\ref{fint}) and define 
the effective CDM density and pressure as Eq.~(\ref{hatrhoc}), 
the energy exchange between CDM and DE induced by the coupling 
$f_1\rho_c$ can be explicitly seen in Eqs.~(\ref{hatrhoceq}) and (\ref{hatPceq}). 
The momentum transfer associated with the coupling $f_2$ does not appear on the 
right-hand-sides of CDM and DE continuity equations at the background level.

In Sec.~\ref{S2sec}, we obtained the second-order action of scalar perturbations 
for the perturbed line element (\ref{permet}) without fixing any particular 
gauge conditions. The resulting full linear perturbation equations of motion are 
given by Eqs.~(\ref{pereq1})-(\ref{eqE2}), which are written 
in the gauge-ready form. 
By introducing several gauge-invariant perturbations in Eq.~(\ref{delphiN}) 
and dimensionless variables in Eq.~(\ref{nodim}), we showed that 
all the perturbation equations are expressed in terms of 
gauge-invariant combinations without residual gauge degrees of freedom, 
see Eqs.~(\ref{perteq1})-(\ref{aniso}). 

In Sec.~\ref{stasec}, we identified stability conditions under which neither
ghost nor Laplacian instabilities are present for scalar perturbations 
deep inside the sound horizon. 
The tensor perturbation does not have a ghost for 
$q_t=2G_4>0$, with the propagation speed $c_t$ equivalent to 
that of light. As long as $q_c$ and $q_s$ given by Eqs.~(\ref{qc}) and 
(\ref{qs}) are positive, the ghosts are absent in the CDM and DE sectors. 
If the coupling $f$ satisfies the condition $f_{,n_c n_c}=0$ with 
$c_c^2=0$, we showed that the effective CDM sound speed squared
$c_{\rm CDM}^2$ vanishes. This includes the interacting theories 
given by the coupling (\ref{fcon2}), for which there are no additional 
pressures preventing or enhancing the gravitational instability of CDM 
perturbations. The mixing between DE and CDM adds a
contribution $\Delta c_s^2$ to the scalar propagation speed squared 
$\hat{c}_s^2$ given by Eq.~(\ref{hatcs}), so the Laplacian instability in the DE sector is 
absent for $c_s^2=\hat{c}_s^2+\Delta c_s^2 \geq 0$.

In Sec.~\ref{Geffsec}, we employed the quasi-static approximation for perturbations 
deep inside the sound horizon to derive the effective gravitational couplings of CDM and baryons 
for the interacting theories satisfying $f_{,n_c n_c}=0$.
The CDM density contrast $\delta_{c{\rm N}}$ obeys the second-order differential 
Eq.~(\ref{ddotceqf}), with $G_{cc}$ and $G_{cb}$ given by  
Eqs.~(\ref{Gcc}) and (\ref{Gcb}) respectively. 
On the other hand, the effective gravitational couplings $G_{bc}$ and $G_{bb}$ 
for baryons are of the forms (\ref{Gbc}) and (\ref{Gbb}). 
Unlike the standard uncoupled Horndeski theories, the growth rate 
$f_c=\dot{\delta}_{c{\rm N}}/(H \delta_{c{\rm N}})$ of CDM perturbations 
appears in the Poisson Eq.~(\ref{Poisson}) and Eq.~(\ref{psiWL}) 
of the weak lensing potential $\psi_{\rm WL}$.

In Sec.~\ref{conthesec}, we applied our general formulas of 
$G_{cc}$, $G_{cb}$,  $G_{bc}$, and $G_{bb}$ for concrete interacting 
theories which belong to the coupling (\ref{fcon3}).
For the theories with $f_1=0$ and $f_2 \neq 0$, the momentum 
exchange between CDM and DE generally gives the values of 
$G_{cc}$ and $G_{cb}$ very different from $G_{bb}$ and $G_{bc}$ 
at late cosmological epochs.
This property is attributed to the fact that the $Z$ dependence in $f_2$
leads to the increase of the quantity $q_c=1+\dot{\phi}f_{2,Z}/\rho_c$.
If the DE sector is described by k-essence or extended Galileons, 
we showed that $G_{cc}$ smaller than $G$ can be naturally realized 
after the dominance of DE. 
The presence of nonvanishing coupling $-f_1(\phi)\rho_c$ besides $f_2$ gives rise 
to additional contributions to $G_{cc}$, but the momentum transfer 
arising from the $Z$ dependence in $f_2$ plays an important role to suppress the CDM 
gravitational couplings at late times. We showed that our general formulas 
of gravitational couplings reproduce the results for specific interacting 
theories known in the literature. 

It will be of interest to apply our general Lagrangian formulation of coupled DE and DM 
to place observational constraints on concrete models.
In particular, the implementation of the perturbation equations of motion 
into the (hi-)CLASS \cite{class,highclass} or EFTCAMB code \cite{EFTCAMB1,EFTCAMB2} 
is the first step for confronting interacting 
models with numerous observational data. 
The perturbation equations derived in this paper are suitable for this purpose, as they are 
written in a gauge-ready/gauge-invariant form
with the EFT-like parameters $\beta_{\rm K}$, $\beta_{n_c}$ besides 
$\alpha_{\rm K}$, $\alpha_{\rm B}$, $\alpha_{\rm M}$. 
We hope that the interacting DE and DM models allow the possibility 
for alleviating the observational tensions of $H_0$ and $\sigma_8$.

\section*{Acknowledgements}

RK is supported by the Grant-in-Aid for Young Scientists B 
of the JSPS No.\,17K14297. 
ST is supported by the Grant-in-Aid for Scientific Research Fund of the JSPS No.\,19K03854 and
MEXT KAKENHI Grant-in-Aid for Scientific Research on Innovative Areas
``Cosmic Acceleration'' (No.\,15H05890).


\end{document}